\documentclass{article}

\usepackage[utf8]{inputenc} %
\usepackage[T1]{fontenc}    %
\usepackage{hyperref}       %
\usepackage{url}            %
\usepackage{booktabs}       %
\usepackage{amsfonts}       %
\usepackage{amssymb}
\usepackage{amsmath}
\usepackage{nicefrac}       %
\usepackage{microtype}      %
\usepackage{lipsum}
\usepackage{todonotes}
\usepackage{cleveref}
\usepackage{graphicx}
\usepackage{subcaption}
\usepackage{tikz}
\usepackage{comment}
\usepackage{natbib}
\graphicspath{{figures/}}
\usetikzlibrary{decorations.pathreplacing}

\def\bbR{\mathbb R}
\def\bbC{\mathbb C}
\def\bbE{\mathbb E}

\def\bx{\mathbf{x}}

\DeclareMathOperator{\Tr}{Tr}
\DeclareMathOperator{\GL}{GL}
\DeclareMathOperator{\KL}{KL}
\DeclareMathOperator{\vc}{vec}

\newcommand{\argmin}{\mathop{\mathrm{arg\,min}}}

\DeclareMathOperator{\Cov}{Cov}
\DeclareMathOperator{\Real}{\mathrm{Re}}

\newcommand\bivec[2]{\begin{bmatrix} #1 \\ #2 \end{bmatrix}}
\newcommand\bimat[4]{\begin{bmatrix} #1 & #2 \\ #3 & #4 \end{bmatrix}}

\title{Spectral independent component analysis with noise modeling for M/EEG source separation}
\author{Pierre Ablin\\ CNRS and DMA, Ecole Normale Supérieure - PSL University\\ Paris, France
\and
Jean-François Cardoso \\ Institut d’Astrophysique de Paris, CNRS (UMR7095)\\ Paris, France
\and
Alexandre Gramfort \\ Inria Saclay, Université Paris-Saclay\\ Palaiseau, France}
\begin{document}

    \maketitle
\begin{abstract}
\emph{Background:}
Independent Component Analysis (ICA) is a widespread tool for exploration and denoising of electroencephalography (EEG) or magnetoencephalography (MEG) signals.
In its most common formulation, ICA assumes that the signal matrix is a noiseless linear mixture of independent sources that are assumed non-Gaussian.
A limitation is that it enforces to estimate as many sources as sensors or to rely on a detrimental PCA step.

\emph{Methods:}
We present the Spectral Matching ICA (SMICA) model. Signals are modelled as a linear mixing of independent sources corrupted by additive noise, where sources and the noise are stationary Gaussian time series.
Thanks to the Gaussian assumption, the negative log-likelihood has a simple expression as a sum of `divergences'  between the empirical spectral covariance matrices of the signals and those predicted by the model.
The model parameters can then be estimated by the expectation-maximization (EM) algorithm.

\emph{Results:}
Experiments on phantom MEG datasets show that SMICA can recover dipole locations more precisely than usual ICA algorithms or Maxwell filtering when the dipole amplitude is low.
Experiments on EEG datasets show that SMICA identifies a source subspace which contains sources that have less pairwise mutual information, and are better explained by the projection of a single dipole on the scalp.

\emph{Comparison with existing methods:} Noiseless ICA models lead to degenerate likelihood when there are fewer sources than sensors,
while SMICA succeeds without resorting to prior dimension reduction.

\emph{Conclusions:}
SMICA is a promising alternative to other noiseless ICA models based
on non-Gaussian assumptions.
\end{abstract}

\section{Introduction}

Magnetoencephalography and Electroencephalography (M/EEG) are popular non-invasive techniques to record brain activity~\cite{hamalainen1993magnetoencephalography,Niedermeyer-da-Silva:05}.
They capture respectively the magnetic and electric signals produced by active neurons from the scalp surface or close to it.
Each M/EEG sensor captures a combination of the different brain activities. The physics of the  mixing is well understood: it is a linear process and can be considered instantaneous.

Independent Component Analysis (ICA)~\cite{hyvarinen2004independent} is extensively used in neuroscience for processing M/EEG signals~\cite{makeig1996independent}.
In its simplest form, it models the observed signals as a linear combination of statistically independent signals called~\emph{sources}.
Remarkably, ICA can identify these sources `blindly', that is, without prior knowledge of the underlying physics of the system (except linearity).
Applied on EEG signals, it separates meaningful brain signals from artifacts (eye blinks, heartbeats, line noise, muscle, $\dots$)~\cite{jung2000removing}, making it an algorithm of choice for artifact rejection~\cite{uriguen2015eeg}.
ICA is widely used for the same purpose in MEG studies~\cite{mantini2008improving,vigario2000independent,ikeda2000independent,dammers-etal:08}.

Beyond artifact removal, ICA is also used to reveal and study brain activity.
In~\cite{makeig2004mining}, ICA is successfully applied to recover evoked and induced event-related dynamics in EEG signals. In~\cite{gomez2008measuring}, ICA is used to extract brain sources, on which causal relations are exhibited, uncovering directional coupling. In~\cite{subasi2010eeg}, independent EEG sources are used in a machine learning pipeline, predicting epileptic seizures.
ICA is used on MEG signals to identify links between function and structure in the brain in~\cite{stephen2013using}. It can also be used on MEG data, coupled with Hilbert filtering, to uncover resting state networks~\cite{brookes2011investigating}.

Finally, ICA components can be mapped to certain brain areas via \emph{source localization}. Indeed, the individual contribution of each source to each sensor can be represented as a topography on the scalp for EEG or on the helmet for MEG. An equivalent current dipole (ECD) can then be fitted to the topography~\cite{scherg-etal:85}, yielding at the same time an estimate of the source location, and its \emph{dipolarity} (how close it can be explained by a focal activity in the brain modeled with a single dipole).
The hypothesis of \emph{independence} of the sources is at the heart of ICA.
However, independence is a statistical property which is difficult to quantify on real data.
In neuroscience, the most widely used algorithms are Infomax~\cite{bell1995information} and FastICA~\cite{hyvarinen1999fast}.
These algorithms perform \emph{non-Gaussian} ICA:
they quantify independence on the marginal (instantaneous) distribution of the data.
They ignore any time correlation and focus entirely on the non-Gaussianity of the data.
In this case, the sources can be recovered when at most one source has a Gaussian density~\cite{comon1994independent}.
Brain sources and artifacts are usually heavy-tail signals which depart form Gaussianity, justifying the use of non-Gaussian algorithms for M/EEG processing.

Another route to ICA is to leverage the time correlation of the sources.
In this case, the sources can be recovered when the sources are \emph{spectrally diverse}, that is, when their power spectrum are non-proportional~\cite{pham1997blind}.
Among these algorithms, Second Order Blind Identification (SOBI)~\cite{belouchrani1997blind} is one of the most widely used.
It jointly diagonalizes a set of time correlation matrices.
Another approach closely related to our work consists in the joint diagonalization of spectral covariance matrices~\cite{pham1997blind}.
ICA methods based on joint-diagonalization of second order statistics might be less popular than non-Gaussian ICA methods, but have encountered some success in M/EEG processing. \citet{congedo2008blind} argues that Pham's approach~\cite{pham1997blind} should be prefered to SOBI for M/EEG preprocessing, one reason being that Pham's approach does not enforce orthogonal constraints.
Finally, in order to leverage both non-Gaussianity and spectral diversity, several methods based on short-time Fourier transform (STFT) have been proposed.
For instance, Fourier-ICA~\cite{hyvarinen2010independent} leverages both non-Gaussianity and spectral diversity with a hybrid method, consisting of the non-Gaussian ICA of concatenated short-time Fourier transforms.
Other approaches consist in joint diagonalization of cospectral matrices (covariance matrices of STFT frames)~\cite{congedo2008blind}
or work in the wavelet domain~\cite{spie03b}.

While all these algorithms rely on various independence measures,
they make the strong assumption that there is no \emph{sensor noise}: they assume that the signal of each sensor is a linear and noiseless combination of sources.
A consequence of the noiseless model is that it enforces that there are as many sources as sensors, while the number of sensors is generally fixed by hardware constraints and not by the actual number of brain or artifactual sources present in the data.
Unfortunately these noiseless models lead to a degenerate likelihood when there are fewer sources than sensors.
This is why, when fewer sources than sensors are expected to be present in the data, a dimension reduction technique like Principal Component Analysis (PCA) is often applied before ICA.
However, this two-stage approach, consisting of first applying PCA and then ICA, is heuristic as based on the assumption that independent sources have high variance, which is not necessary.
As it is argued in~\cite{artoni2018applying} applying PCA before ICA can degrade the quality of the recovered sources. To avoid relying on PCA from dimensionality reduction, it is also sometimes suggested to simply discard some channels. Throwing away data without a clear motivation is arguably questionable.

In order to alleviate this problem, some ICA algorithms incorporate a noise model.
As explained in ~\cite{hyvarinen1998independent}, when the noise statistics are known, maximizing the likelihood of such a model is an optimization problem sharing many similarities with dictionary learning~\citep{olshausen1996emergence}.
Such procedure is typically much more costly than regular ICA, and it is seldom used in M/EEG processing (See ~\cite{barthelemy2013multivariate, liu2017sparse} for instance).
In~\citep{parra2000convolutive}, noise is modelled in a non-stationary framework: the sources -and noise- are assumed non-stationary (their instantaneous variance varies over time). The model is estimated by fitting it to the data, via the minimization of a simple quadratic criterion, which deviates from the probabilistic model.

In this article, we study Spectral Matching ICA (SMICA) for M/EEG processing.
This ICA model has been first investigated in astronomy for separation of the cosmic microwave background~\cite{cardoso2002blind, delabrouille2003multidetector}.
SMICA models the observations as a sum of a linear mixture of independent sources and noise. It assumes that the sources and noise are Gaussian, and that the sources have non-proportional spectra.
This assumption makes it well suited for brain rhythms and artifacts extraction as they are known to have prototypical spectra. Brain sources tend to exhibit so-called ``1/f'' power spectral densities, while the spectra of artifacts are often localized in certain frequency bands (e.g. muscle artifacts or line noise).
Importantly, the statistics of the noise are parameters of the model, and are estimated along the other parameters of the model.
Thanks to its noise model, SMICA can estimate fewer sources than sensors.
Therefore, no preprocessing for dimension reduction is required.
The sources can be estimated by Wiener filtering, which takes the noise estimation into account and denoises the sources.

The article is organized as follows.
In \autoref{sec:smica}, the SMICA statistical model is introduced and the estimation strategy based on an Expectation-Maximisation (EM) algorithm is described.
In \autoref{sec:expe}, the usefulness of SMICA is demonstrated on various MEG and EEG datasets.
\smallskip

\textbf{Notation}
The trace of a matrix $M\in\bbR^{p\times p}$ is $\Tr(M)$, and its determinant is $|M|$. A matrix is invertible when $|M| \neq 0$, and we write $M \in \GL_p$. Given a vector $u \in \bbR^p$, the matrix $\text{diag}(u) \in \bbR^{p\times p}$ is the matrix containing the elements of $u$ on its diagonal, and $0$ elsewhere.
If $M$ is a $p \times p$ matrix, then $\text{diag}(M)$ is the diagonal matrix with the same diagonal as $M$.
Given $A\in \bbR^{p\times q}$, the vectorization of $A$ is a vector $\vc(A) \in \bbR^{pq}$ of entries $\vc(A)_{i+p(j-1)}=A_{ij}$.
The Moore-Penrose Pseudo-Inverse of a tall matrix $A\in\bbR^{p\times q}$ is $A^{\dagger}=(A^\top A)^{-1}A^\top$.

\section{A maximum  likelihood approach to noisy ICA}\label{sec:smica}

\def\bbZ{\mathbb Z}
\def\inv{^{-1}}
\newcommand\floor[1]{\lfloor #1 \rfloor}

This section introduces our approach to blind source separation for noisy observations
(Sec.~\ref{sec:datamodelspec}).  Its application is then discussed in detail
(Sec.~\ref{sec:smica_in_depth}).

\subsection{The SMICA method in theory}\label{sec:datamodelspec}

In a noisy ICA model, the outputs of $p$ sensors, \textit{e.g.} M/EEG recordings,
collected in a vector $X(t)\in \bbR^p$, are modelled as noisy instantaneous mixtures of
$q$ independent sources represented by a vector $S(t)$ of size $q$ with an
additive noise term $N(t)$ of size $p$, this is
\begin{equation}
    \label{eq:ica_plus_noise}
    X(t) = AS(t)+N(t),
\end{equation}
where $A$ is the $p\times q$ mixing matrix.
The noise is assumed independent from the sources and uncorrelated across sensors.

This model readily translates into the spectral domain.
Recall that for a zero-mean $p$-dimensional stationary time series, $\{ X(t) \}$, the
$p\times p$ autocovariance matrix $\bbE[ X(t)X(t+\tau)^\top]$ does not depend on $t$ and
that its Fourier transform%
\footnote{For simplicity we have set the sampling period to one time unit.}:
\begin{equation}\label{eq:defspec}
  \textstyle
  C(f) = \sum_\tau \bbE[ X(t)X(t+\tau)^\top] \,e^{-2i\pi f\tau}
\end{equation}
defines $p \times p$ {\it spectral covariance matrices} $C(f)$.
The diagonal entry $C_{aa}(f)$ is the power spectrum of $\{X_a(t)\}$ while $C_{ab}(f)$
contains the cross-spectrum between $\{X_a(t)\}$ and $\{X_b(t)\}$.

The linear relation between data and sources of Eq.~(\ref{eq:ica_plus_noise}) translates
into the spectral model
\begin{equation}
  \label{eq:SCMX}
  C(f) = A P(f) A^\top + \Sigma(f)
\end{equation}
where $P(f)$ and $\Sigma(f)$ are the spectral covariance matrices of sources and of the
noise. In this work, we assume that the sources and the noise terms are independent, which means that $P(f)$ and $\Sigma(f)$ are \textbf{diagonal} matrices.

This particular structure of the spectral covariance matrices is preserved when spectra
are averaged over frequency bands.
Define $B$ frequency intervals $I_1,\ldots,I_B$ by
$I_b = [ f_\mathrm{min}^b , \, f_\mathrm{max}^b ] $ and consider frequency averages over
those bands:
\begin{equation}\label{eq:defavg}
  C_b
  =
  \frac1{f_\mathrm{max}^b - f_\mathrm{min}^b } \int_{f_\mathrm{min}^b}^{f_\mathrm{max}^b}  C_X(f) df
\end{equation}
Then, upon averaging, Eq.~(\ref{eq:SCMX}) becomes
\begin{equation}
  \label{eq:SCMbin}
  C_b  = A P_b A^\top + \Sigma_b
\end{equation}
where $P_b$ and $\Sigma_b$ denote the corresponding averages for $P(f)$ and $\Sigma(f)$.
As a consequence, the noisy ICA model in Eq.~\eqref{eq:ica_plus_noise} is transformed in the simpler model of Eq.~\eqref{eq:SCMbin}, where the parameters are the mixing matrix $A$, the source powers in each band $P_b$, and the noise powers in each band, $\Sigma_b$.

The noisy ICA model is inferred from by connecting the spectral matrices $C_b$ of
model~(\ref{eq:SCMbin}) to samples estimates.  If $T$ data samples $X(0),\ldots, X(T-1)$
are avialble, spectral matrices are classically estimated from the Fourier coefficients
\begin{equation}
  \label{eq:fourier}
  \tilde{\bx}_k=\frac1{\sqrt{T}} \sum_{t=0}^{T-1} X(t)\,  e^{-2i\pi kt/T}
\end{equation}
by averaging over the relevant frequency bands.  These estimates are:
\begin{equation}
  \label{eq:defempspec}
  \widehat C _b = \frac1{n_b} \sum_{k: \frac k T \in I_b}  \Real\bigl( \tilde{\bx}_k\tilde{\bx}_k^H \bigr)
\end{equation}
where $ n_b = \# \{k: \frac k T \in I_b \} $ denotes the number of Fourier coefficients
available in band $b$.

\begin{figure}
  \centering
    \begin{tikzpicture}
      \node (img1) {\includegraphics[height=1.5cm]{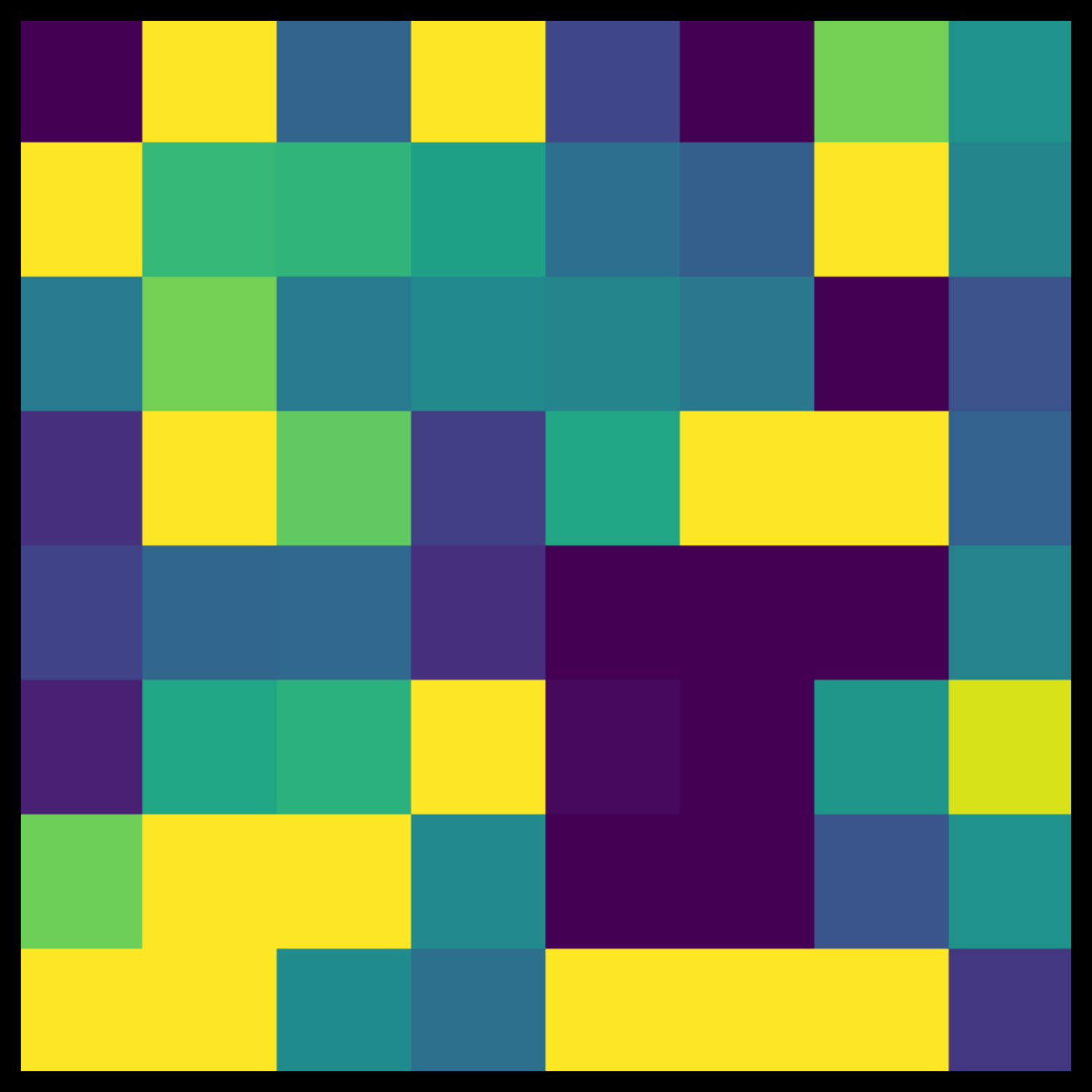}};
      \node (img2) at (img1.south east) [xshift=-.5cm, yshift=-.2cm] {\includegraphics[height=1.5cm]{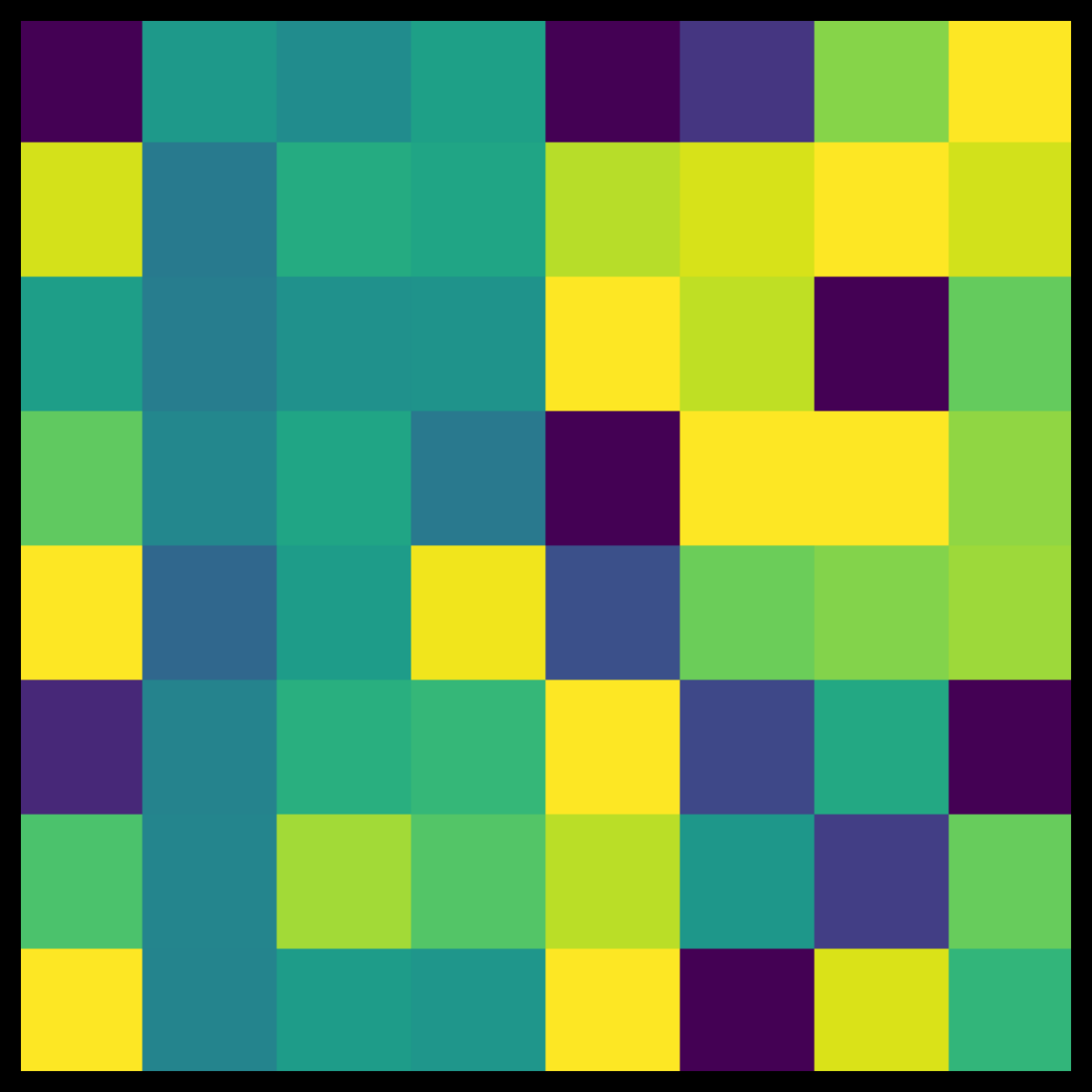}};
      \node (img3) at (img2.south east) [xshift=-.5cm, yshift=-.2cm] {\includegraphics[height=1.5cm]{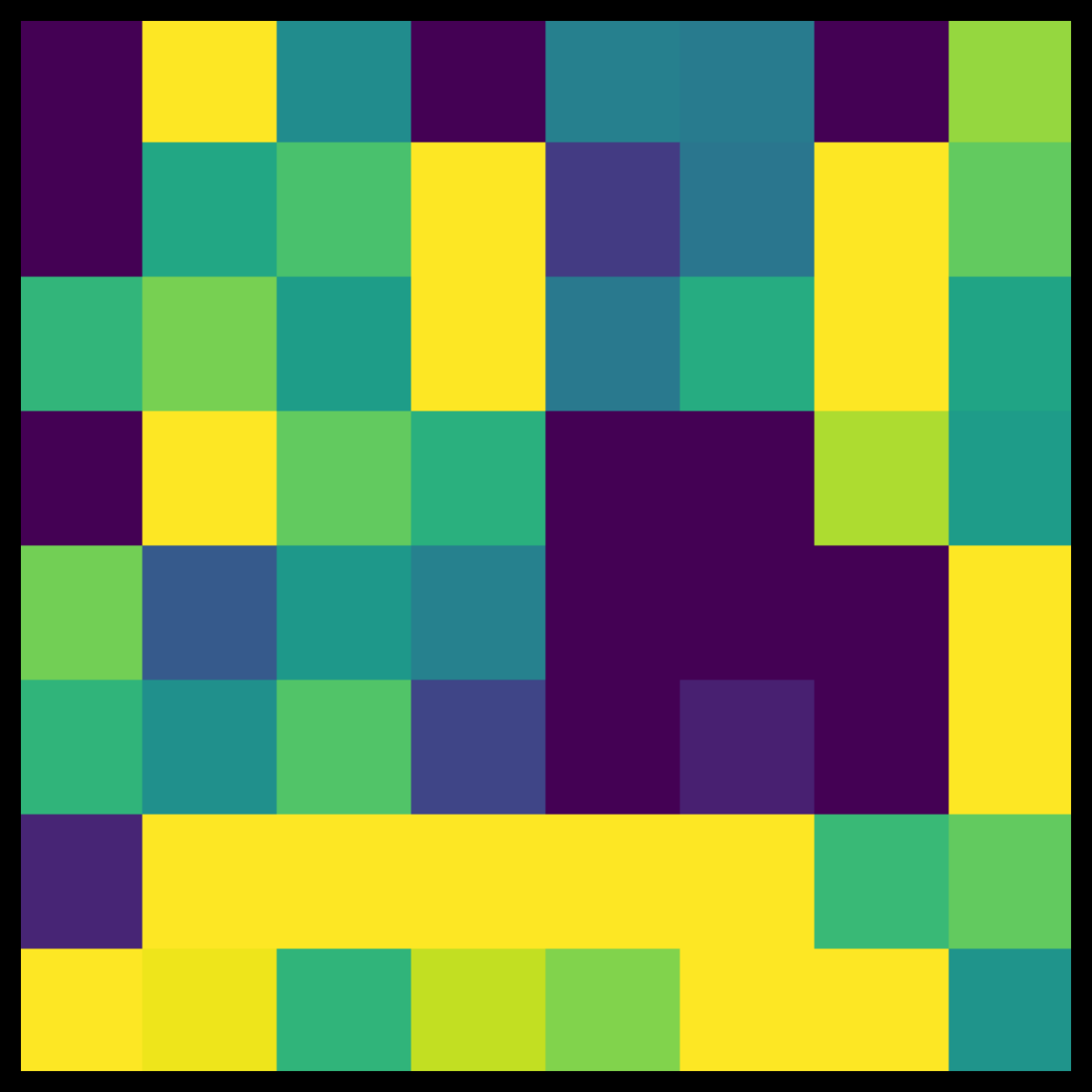}};
      \node (eq) at (img2.east) [xshift=.4cm]{$\simeq$};
      \node (C1) at (img1.west) [xshift=-.2cm, yshift=-.1cm] {$\hat{C}_1$};
      \node (C2) at (img2.west) [xshift=-.2cm, yshift=-.1cm] {$\hat{C}_2$};
      \node (CB) at (img3.west) [xshift=-.2cm, yshift=-.1cm] {$\hat{C}_B$};
      \draw[dashed] (C2) -- (CB);
      \node (A) at (eq.west) [xshift=1cm] {\includegraphics[height=1.5cm]{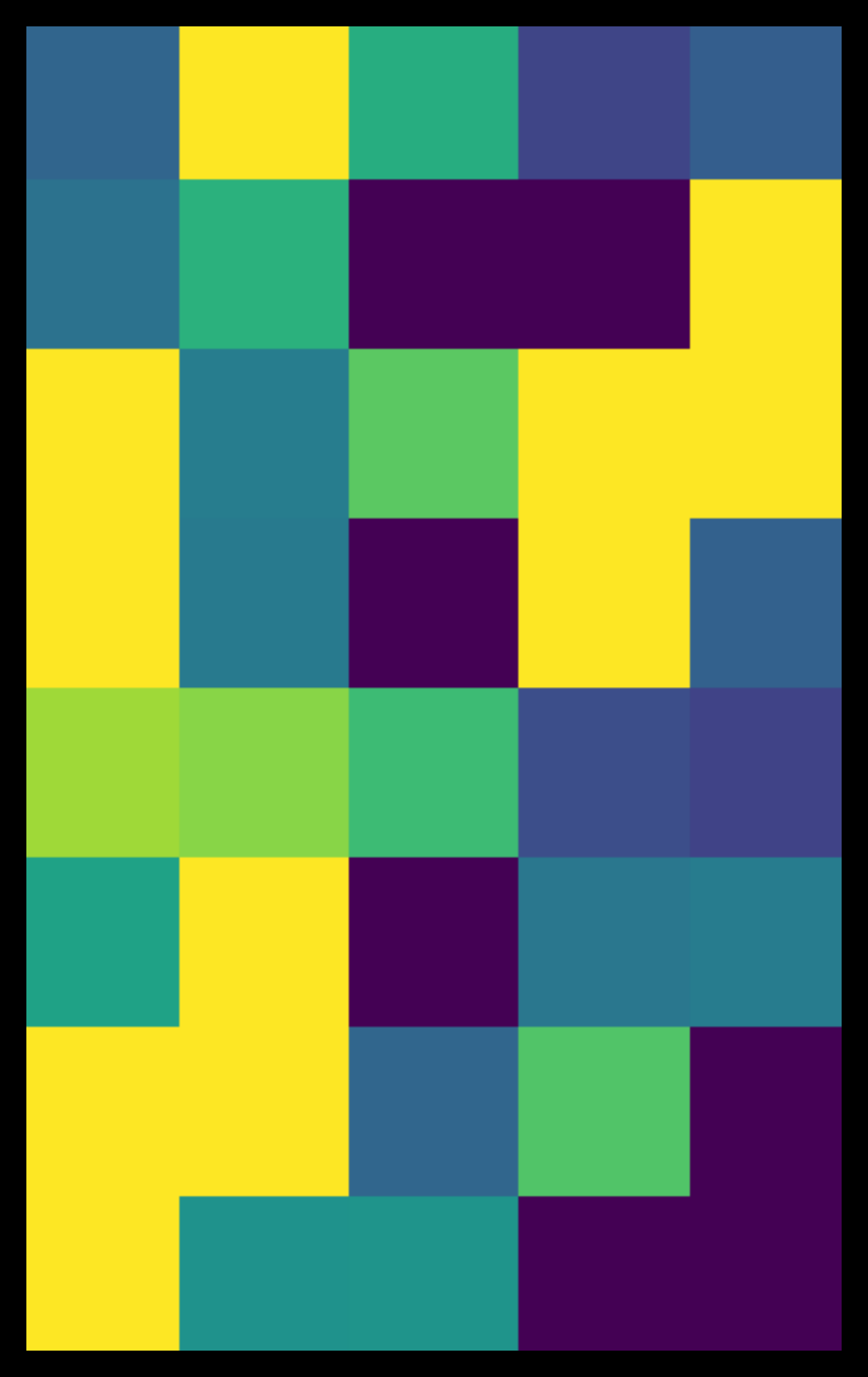}};
      \node (name) at (A.north) [yshift=.2cm] {$A$};
      \node (pimg1) at (img1.east) [xshift=3.2cm] {\includegraphics[height=0.9cm]{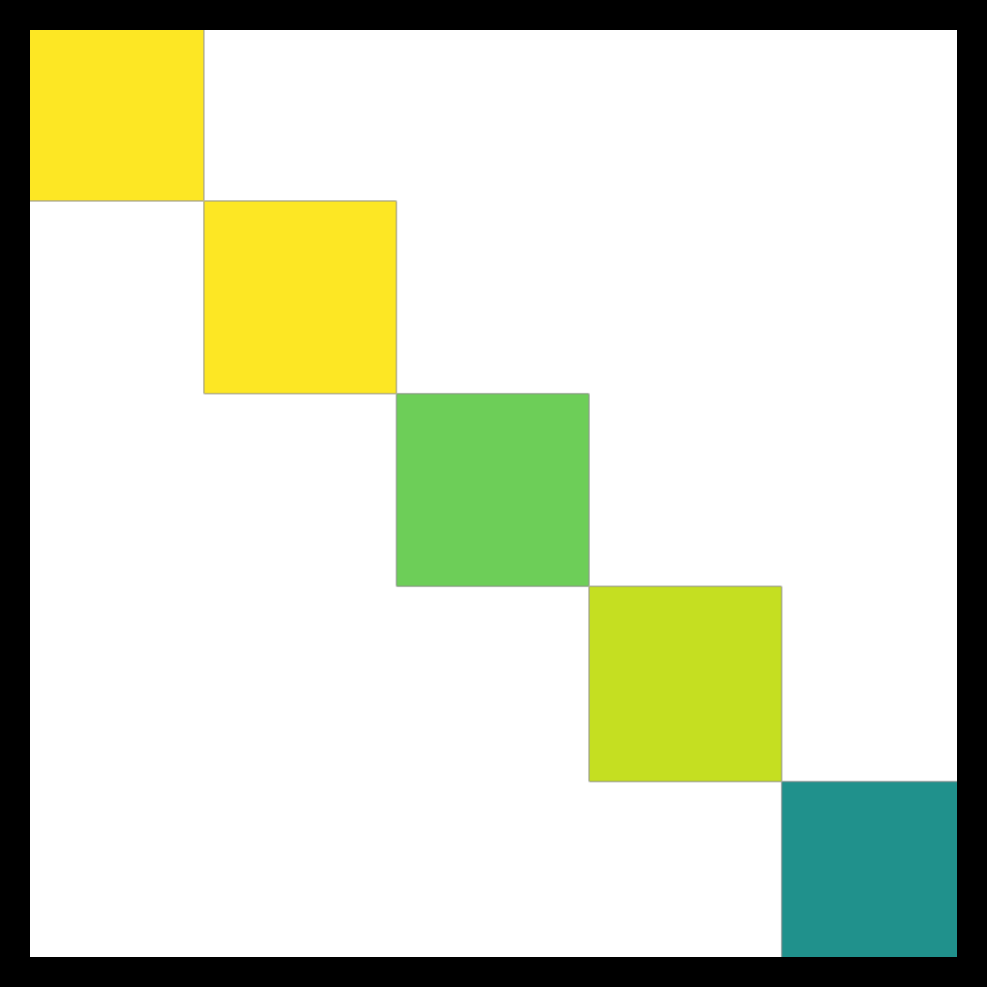}};
      \node (pimg2) at (pimg1.south east) [xshift=-.4cm, yshift=-.47cm] {\includegraphics[height=0.9cm]{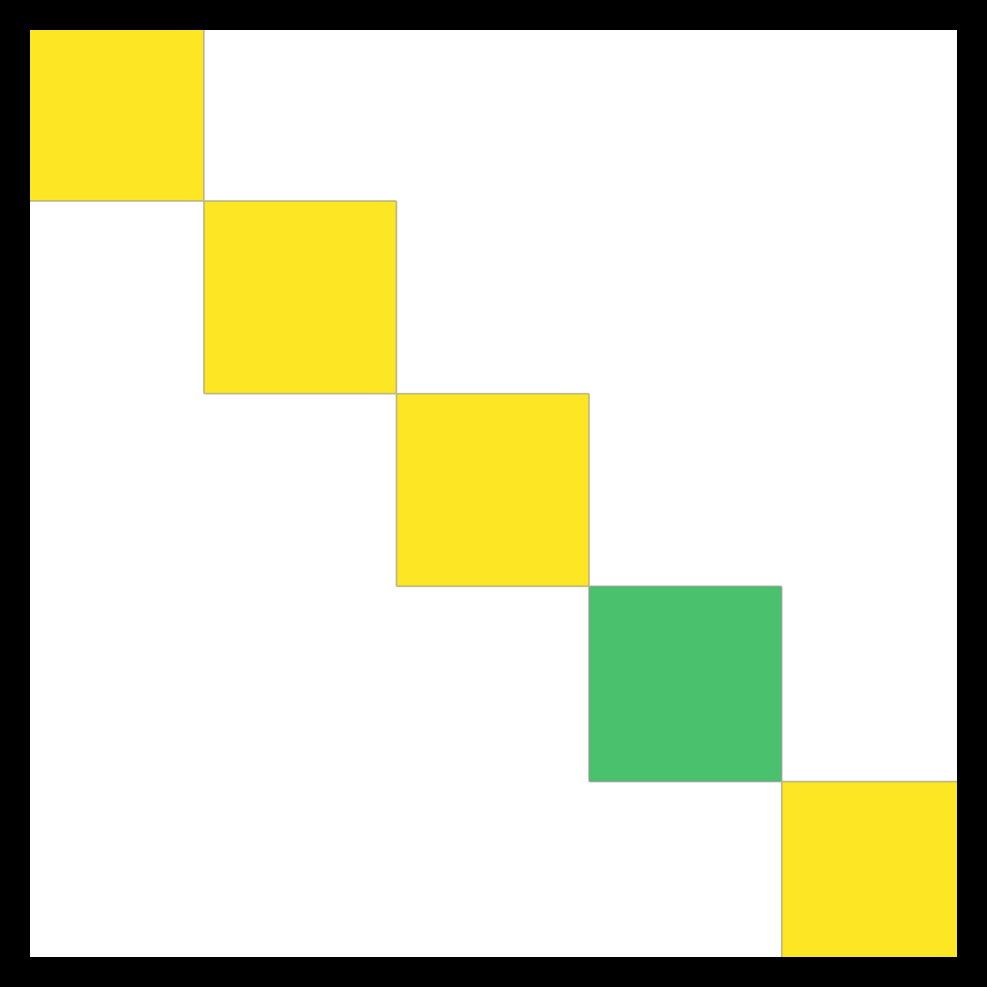}};
      \node (pimg3) at (pimg2.south east) [xshift=-.4cm, yshift=-.47cm] {\includegraphics[height=0.9cm]{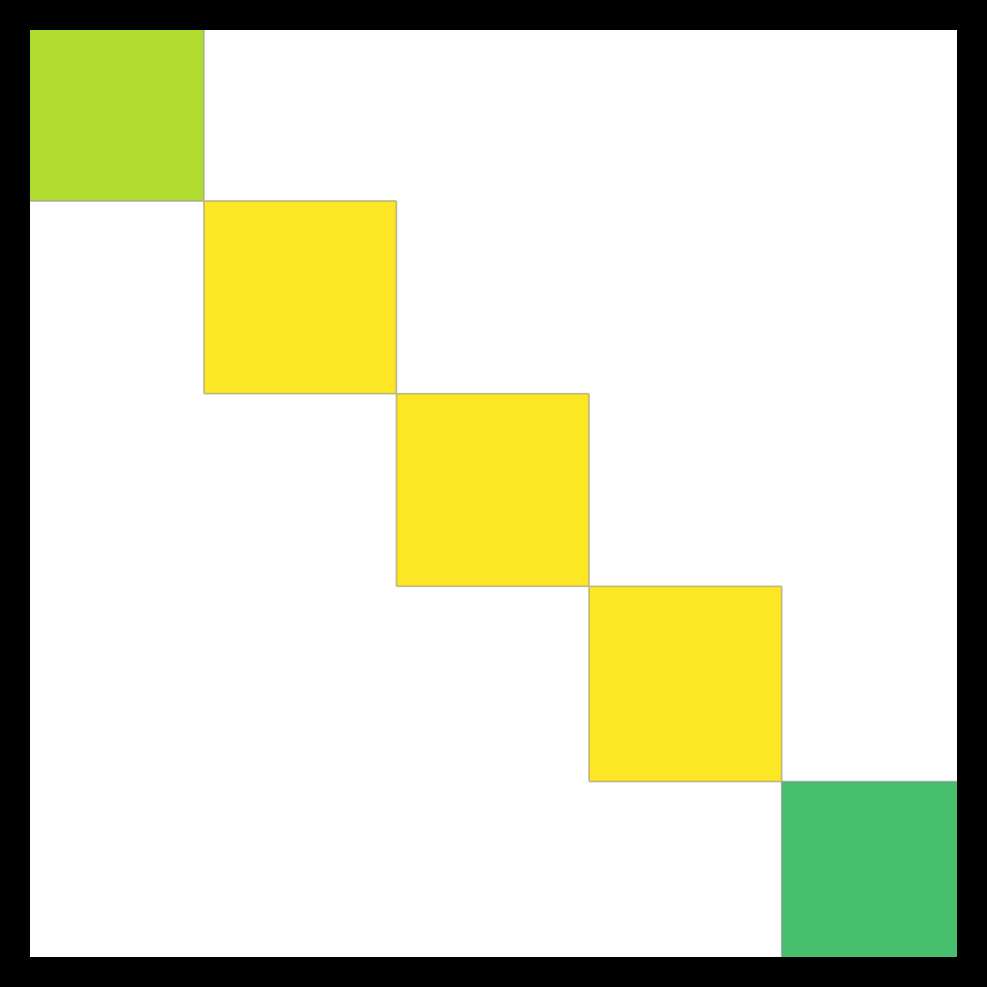}};
      \node (AT) at (A.west) [xshift=4.1cm] {\includegraphics[height=.9cm]{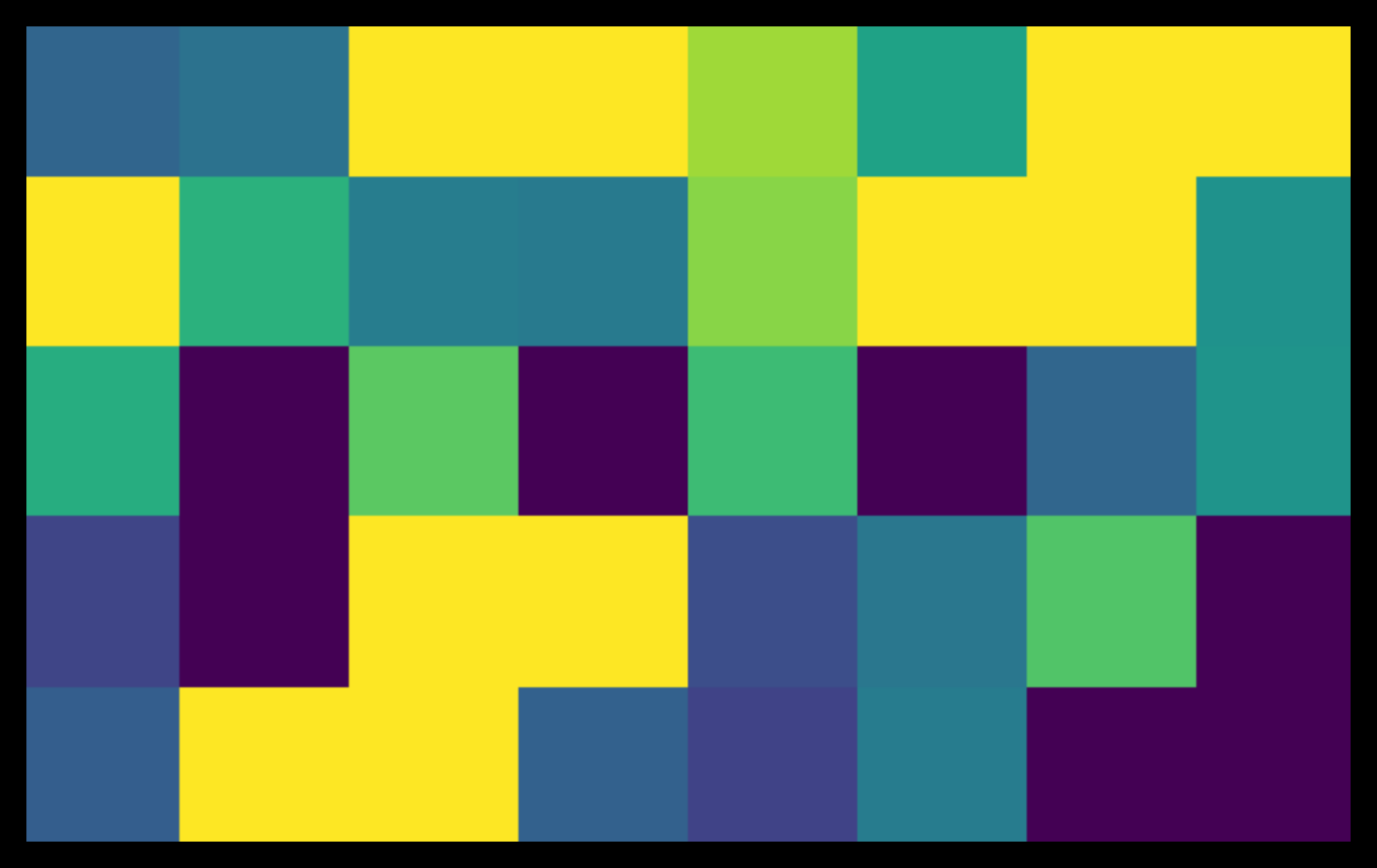}};
      \node (nameT) at (AT.north) [yshift=.2cm] {$A^{\top}$};
      \node (plus) at (AT.east) [xshift=.3cm]{$+$};
      \node (nimg1) at (img1.east) [xshift=7.7cm] {\includegraphics[height=1.5cm]{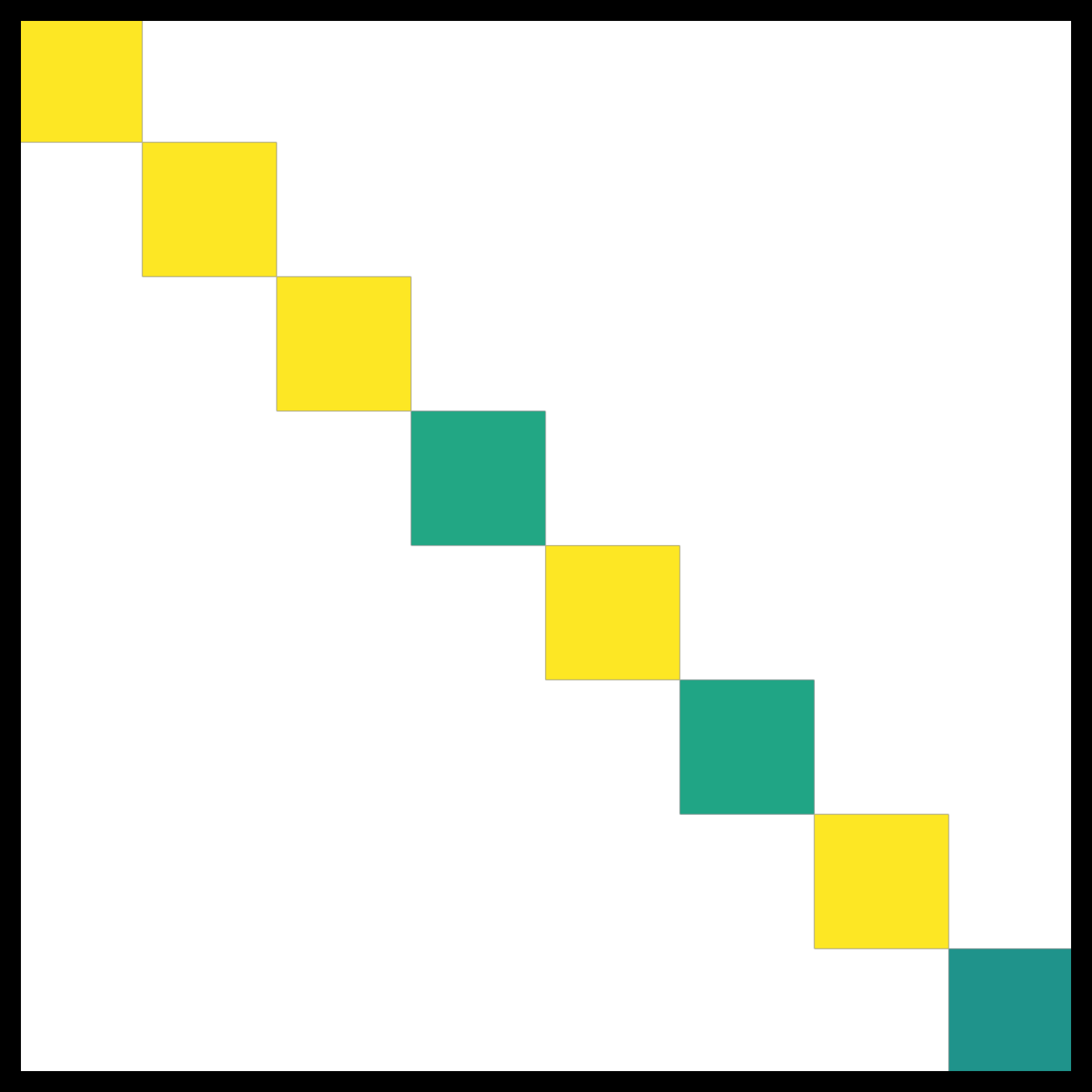}};
      \node (nimg2) at (nimg1.south east) [xshift=-.5cm, yshift=-.2cm] {\includegraphics[height=1.5cm]{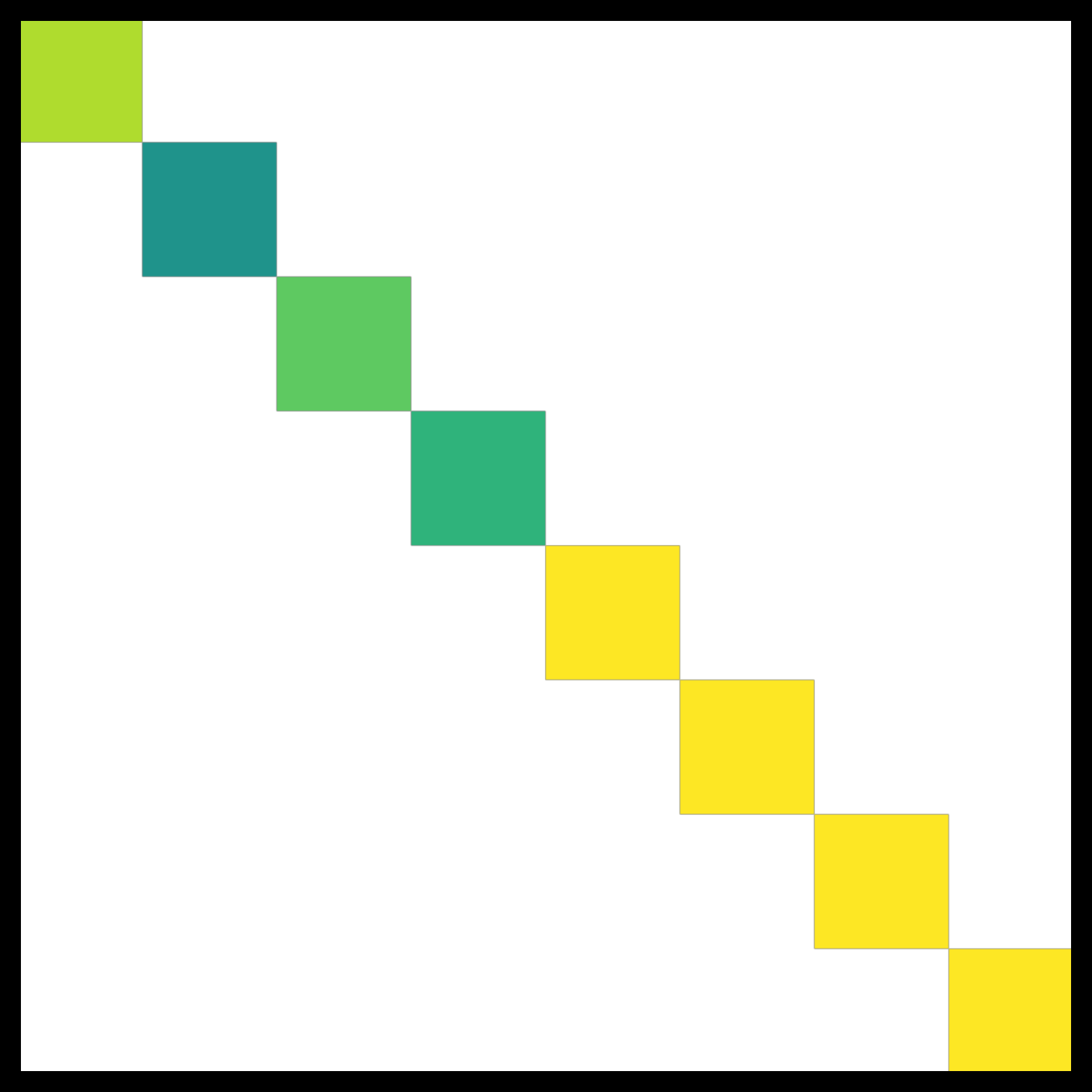}};
      \node (nimg3) at (nimg2.south east) [xshift=-.5cm, yshift=-.2cm] {\includegraphics[height=1.5cm]{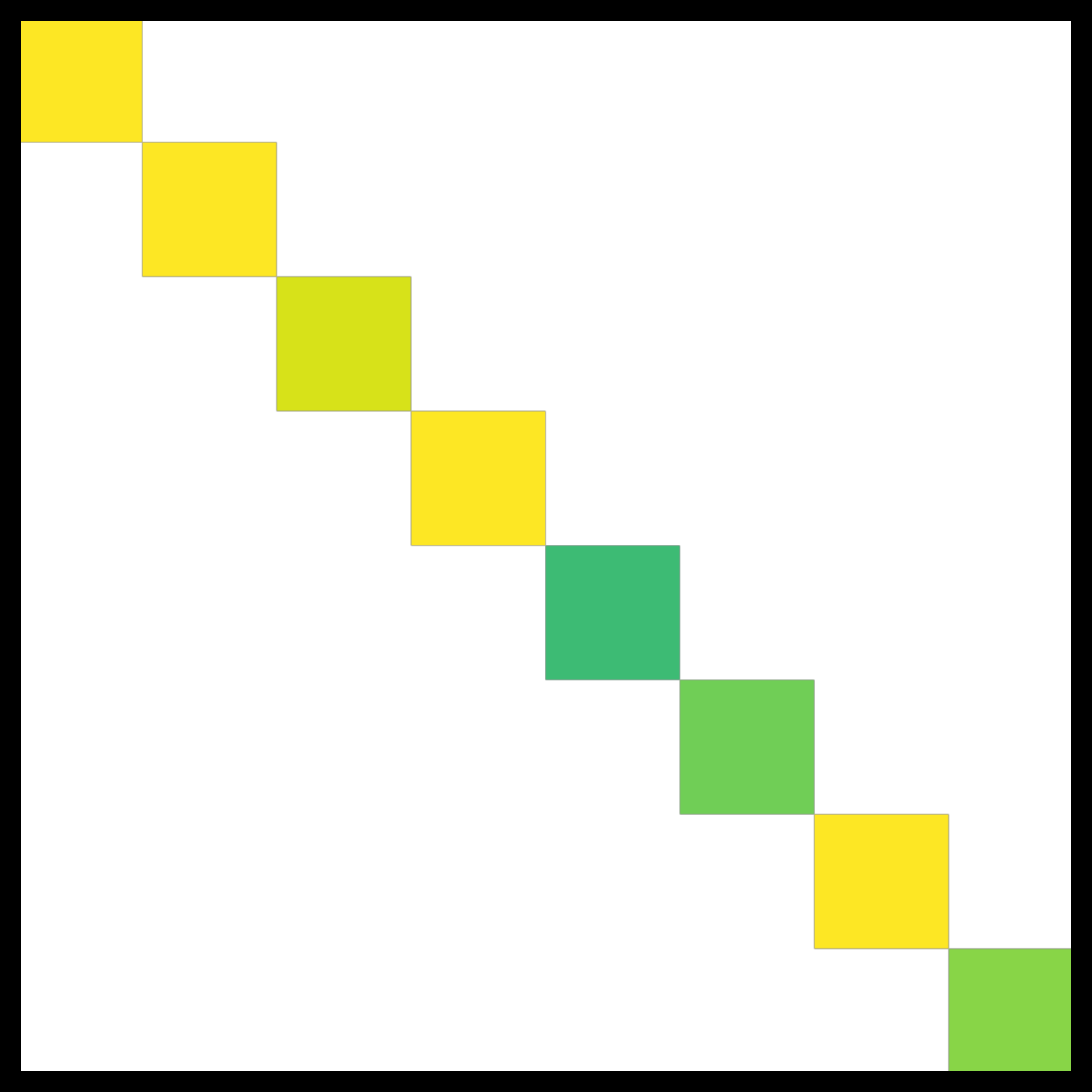}};
      \node (S1) at (nimg1.west) [xshift=-.2cm, yshift=-.1cm] {$\Sigma_1$};
      \node (S2) at (nimg2.west) [xshift=-.2cm, yshift=-.1cm] {$\Sigma_2$};
      \node (SB) at (nimg3.west) [xshift=-.2cm, yshift=-.1cm] {$\Sigma_B$};
      \draw[dashed] (S2) -- (SB);
      \node (P1) at (pimg1.west) [xshift=-.15cm, yshift=-.1cm] {$P_1$};
      \node (P2) at (pimg2.west) [xshift=-.15cm, yshift=-.1cm] {$P_2$};
      \node (PB) at (pimg3.west) [xshift=-.15cm, yshift=-.1cm] {$P_B$};
      \draw[dashed] (P2) -- (PB);
     \draw [decorate,decoration={brace,amplitude=5pt}] (img1.north west) -- (img1.north east) node [black,midway, yshift=.4cm] {Sensors};
      \draw [decorate,decoration={brace,amplitude=5pt}] (pimg1.north west) -- (pimg1.north east) node [black,midway, yshift=.4cm] {Sources};
    \end{tikzpicture}
    \caption{The SMICA method: empirical spectral covariance matrices
      $\hat{C}_1, \dots, \hat{C}_B$ computed from the M/EEG data are matched by the model
      $\hat{C}_b \simeq A P_b A^{\top} + \Sigma_b$, where the $P_b$ and $\Sigma_b$ are
      diagonal positive matrices.  Matching is performed in a statistically sound way, by
      using a matching criterion derived from a simple likelihood.}
    \label{fig:smica_principle}
\end{figure}
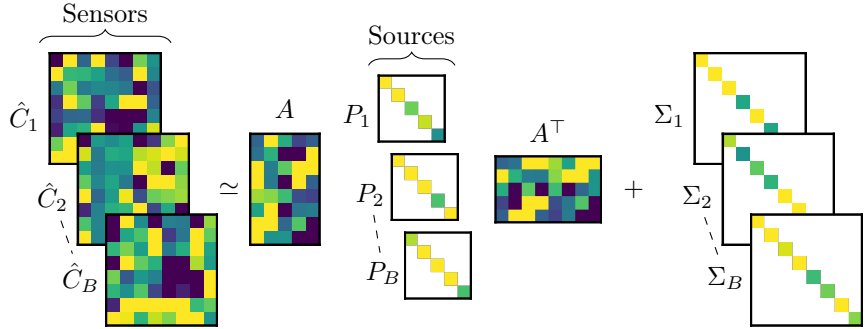

The set $\theta=\left(A, P_1, \ldots, P_B, \Sigma_1,\ldots,\Sigma_B\right)$ of all unknown
parameters can be estimated by adjusting the model $C_b = A P_b A^\top + \Sigma_b$ to the
data as summarized by $\widehat C _b$.
This spectral matching principle is illustrated by Figure~\ref{fig:smica_principle}.
We advocate using a specific spectral matching criterion:
\begin{equation}
    \label{eq:gaussian_likelihood}
    \mathcal{L}(\theta)
    =
    \sum_{b=1}^B  2 n_b \KL\left(\widehat{C}_b, \, A P_bA^\top + \Sigma_b \right)
\end{equation}
where $\KL$ is the Kullback-Leibler divergence between two $p\times p$ positive matrices:
\begin{equation}
  \label{eq:kl}
  \KL(C_1, C_2)
  =
  \frac12 \Bigl( \Tr(C_1C_2^{-1}) -\log\det(C_1C_2^{-1}) - p \Bigr) .
\end{equation}
The KL-divergence $\KL(C_1, C_2)$ is non-negative and cancels if and only if $C_1=C_2$.

The particular measure~(\ref{eq:gaussian_likelihood}) of spectral adjustment between data
and model has been chosen because it is (up to an irrelevant constant) asymptotically (for
large $T$ and narrow bands) equal to minus the log likelihood of a Gaussian stationary
model.
Hence, the SMICA estimates inherits some good properties of maximum likelihood estimates,
in particular they enjoy a built-in scale invariance and they can be easily computed using
the EM algorithm.
Those properties and other considerations are discussed in the remaining of the section. The derivation of the SMICA criterion from a likelihood function is explained in~\ref{sec:SMLL}.

\subsection{Statistical properties of SMICA}
\label{sec:smica_in_depth}

The novelty in SMICA comes from its noise model: ICA methods that work in the spectral domain are common. In fact, one recovers a  popular method by removing the noise in Eq.~\eqref{eq:SCMbin}, and by assuming a square mixing
matrix $p=q$, that is, finding as many sources as sensors.
This was considered by Pham who used the Whittle approximation in his seminal
paper~\cite{pham1997blind}.
In that case ($p=q$, $N(f)=0$), the spectral mismatch can be shown to reduce to a joint-diagonality criterion.  Estimating $A$ amounts to solving the~\emph{joint-diagonalization}
of the spectral covariances: $A$ is such that the matrices $A^{-1}\widehat{C}_bA^{-\top}$
are as diagonal as possible.  The use of such approach for EEG processing is advocated
in~\citep{congedo2008blind}.
In the experimental section~\ref{sec:expe}, we compare SMICA to this approach of a plain
joint diagonalization of spectral matrices and we refer to it as \emph{JDIAG}.

The noise model in SMICA makes the estimation of the model parameters harder than joint-diagonalization, but it enables finer source estimation, through \emph{Wiener filtering}  (\citep{brown1992introduction}, Chapter 4).
Noiseless ICA models of the form $X = AS$ have a simple way of computing the estimated sources: $\widehat{S} = A^{-1}X$.
SMICA can employ the same technique for recovering the source, albeit replacing $A^{-1}$ by $A^{\dagger}$, the Moore-Penrose pseudo-inverse of $A$. However, ) it is more appealing to compute the expected value of the sources given the parameters: $\widehat{S} = \mathbb{E}[S | X, \theta]$, which is given by Wiener filtering in each frequency band:
\begin{equation}
    \label{eq:wiener}
    \widehat{S}_b = (A^{\top} \Sigma_b^{-1}A + P_b^{-1})^{-1}A^{\top} \Sigma_b^{-1}X_b \enspace ,
\end{equation}
where $X_b$ is the matrix of signals filtered in the frequency band $[f_b, f_{b+1}]$.
This operation is linear in $X$, and is adaptive to the level of noise: if in frequency band $b$ the estimated noise $\Sigma_b$ for a sensor is large, then its contribution in the source estimate shrinks towards $0$.
Note that the standard ICA source estimation formula is recovered when the noise is equal on all channels and tends to $0$ (i.e. $\Sigma_b = \lambda_b I_p$ with $\lambda_b \to 0$), yielding at the limit $\widehat{S} = A^{\dagger} X$.

Next, we study the \emph{identifiability} conditions for SMICA.
The issue of identifiability is to find the conditions allowing for a unique recovery of
the sources.  Identifiability conditions of \emph{noiseless} spectral ICA are well
established~\cite{pham1997blind}: the sources should not have proportional spectra, that
is, for any pair $(i,j)$ of sources the quantities $[P_b]_{ii} / [P_b]_{jj}$ should not be
constant with respect to $b=1,\ldots,B$.
The identifiability conditions for the SMICA model~\eqref{eq:ica_plus_noise} are more
complicated, and the non-proportionality of the source spectra is only a \emph{necessary}:
SMICA can only separate sources with different `spectral signatures'.
Hence, it is well suited to isolate all kinds of artifacts from brain activity: line noise, muscular activity, heartbeats, eye blinks, since their spectra are usually very different from those of brain activity.
Among brain sources, some might have similar power spectrum, such as occipital dominant alpha rhythms and therefore cannot be easily separated by SMICA.

Finally, we discuss the \emph{expectation-maximisation} algorithm to estimate the parameters. Since it corresponds to a likelihood function, the spectral
mismatch~(\ref{eq:gaussian_likelihood}) can be minimized using the
Expectation-Maximization (EM) algorithm~\cite{dempster1977maximum}, where the source and
noise signals are the latent (unobserved) variables.
The EM algorithm is appealing because it does not require any hyperparameters like
learning rates, and is guaranteed to decrease the loss function at each iteration.
Still, this approach is generally slow (it might require many iterations to reach a
satisfactory set of parameters), and other optimization techniques could be investigated
for the fast minimization of the negative log-likelihood $\mathcal{L}$.
The EM algorithm for SMICA is described in the appendix (section~\ref{sec:em_smica}).

Next, we discuss the practical application of SMICA for M/EEG processing.

\subsection{SMICA for M/EEG processing}
\label{sec:smica_in_depth}

This section goes through various considerations regarding the application of SMICA to
M/EEG signals.
A first advantage of SMICA over other ICA algorithms for M/EEG processing is its embedded dimension reduction.
It is often the case that there are more sensors than sources which can be significantly
recovered from the observations.  When $p>q$ matrix $A$ cannot be inverted and one
usually performs some dimension reduction in order to apply ICA algorithms which require a
square matrix ($p=q$) to operate.
In SMICA, thanks to the presence of a noise term, spectral covariance matrices are not
degenerate even if $q<p$ and a non-square matrix $A$ can be fitted by SMICA in a
statistically sound way even when there are fewer sources than sensors, without resorting
to a pre-processing stage of dimension reduction.  In some sense, one can say that SMICA
has a \emph{built-in dimension reduction} of the signal part because it can fit a tall ($p<q$)
mixing matrix.
It often argued that reduction dimension in EEG processing deteriorates the quality of the subsequent ICA decomposition~\cite{artoni2018applying}.
SMICA offers a simple way to circumvent this problem, by embedding dimension reduction in the ICA in a statistically sound way.

\paragraph{Signal denoising using SMICA}

Like any ICA algorithm, SMICA estimates sources which can be marked as spurious / non-biological by specialists: in addition to brain sources, it usually recovers physiological artifacts (heartbeats, eye blinks) and external electromagnetic perturbations (room and line noise).
The remaining sources can then be projected back in the signal space, giving clean M/EEG signals.
Thanks to its noise modeling, SMICA makes these two operations statistically sound: the sources $\widehat{S}$ are estimated by Wiener filtering, some sources can be manually or automatically marked as spurious. If the source $i$ is marked as spurious, we set $\widehat{S}_i = 0$. The cleaned M/EEG signals are computed as $X_{cleaned} = A \widehat{S}$.

\paragraph{Combining SMICA with another ICA algorithm}

In practice, it might happen that some sources recovered by SMICA have similar spectra, which indicates that these sources are not well separated.
Another ICA procedure (typically, based on non-Gaussianity) can then be applied on these sources in order to better separate them.

Taking it a step further,
SMICA can also be used as a \emph{source subspace} identifier. Applied on the M/EEG signals $X \in \bbR^{p\times T}$, SMICA produces a mixing matrix $A_1 \in \bbR^{p \times q}$ and a source matrix $S_1 \in \bbR^{q \times T}$ estimated by Wiener filtering.
The sources in $S_1$ are maximally independent with respect to the separation criterion of SMICA, but might not be maximally independent with respect to another criterion.
Applying another ICA algorithm on $S_1$ yields a square mixing matrix $A_2 \in\bbR^{q \times q}$, and a new source matrix $S_2 = A_2^{-1}S_1$.
The overall mixing matrix linking the sources $S_2$ and the original dataset $X$ is the matrix product $A_{Total} =A_{1} A_{2}$.
For instance, using a non-Gaussian ICA algorithm like Infomax or FastICA on the sources found by SMICA may disentangle sources that share similar power spectrum.
The practical benefits of such approach are demonstrated in the experiments presented in section~\ref{sec:expe}.

\paragraph{Complexity of SMICA}

SMICA only needs the spectral covariances $\hat{C}_1, \dots, \hat{C}_B$ to infer the parameters of the model.
Therefore, the complexity of the EM algorithm does not depend on the number of samples $T$; only the computation of the covariances does.
This differs from non-Gaussian ICA algorithms, for which the estimation time is proportional to the length of the recordings.

In practice, we found that fitting SMICA on a $102$ sensors MEG dataset with $40$ frequency bins and $100$ sources takes about $15$ minutes using one CPU of a recent laptop.

\paragraph{Frequency selection}

SMICA exploits spectral information, but not necesarily over the whole frequency range.
Typically, one may exclude the highest frequencies if they are dominated by noise or cut
off by a sampling filter.  One may also ignore the very lowest frequencies if they are
dominated by slow drift artifacts.  In general, there is no counter indication to
restricting the sum~(\ref{eq:gaussian_likelihood}) to the high SNR part of the frequency
range.

\section{Experiments}
\label{sec:expe}

We report some experiments comparing our approach to other (noiseless) ICA algorithms: these algorithms all model the dataset $X$ as $X = AS$, where $A \in \bbR^{p\times p}$ is the \emph{square} mixing matrix, and $S$ is the source matrix.
However, they estimate the independent components based on different independence criterion.

We start by briefly describing the ICA algorithms which are compared to SMICA.

\paragraph{Non-Gaussian ICA}

Non-Gaussian ICA algorithms model the source time-series as independent and identically distributed, with non-Gaussian probability density functions.
Some of the most popular ICA algorithms fall in this category: FastICA~\cite{hyvarinen1999fast}, Infomax~\cite{bell1995information}, its extended version~\cite{lee1999independent}, JADE~\cite{cardoso1993blind} and more recently AMICA~\cite{palmer2012amica}.

\paragraph{Second order blind identification}

The SOBI algorithm~\cite{belouchrani1997blind} aims at recovering sources with spectral diversity, just like SMICA and JDIAG.
It does so in a heuristic way, by joint-diagonalization of a set of correlation matrices $\frac{1}{T - \tau}\sum_{t=1}^{T-\tau}X(t)X(t - \tau)^{\top}$ for a set of time lags $\tau_1,\dots,\tau_B$, rather than spectral covariance matrices.
Choosing an appropriate set of time lags is not an obvious task; we use the set advised in~\cite{tang2005validation}.
Unlike JDIAG, the joint diagonalization criterion is ad-hoc, and does not correspond to a principled statistical criterion.
For instance, since JDIAG follows the maximum-likelihood principle, it is asymptotically Fisher-efficient and reaches the Cramer-Rao lower bound, unlike SOBI.
This is why several articles argue for the use of JDIAG rather than SOBI~\cite{doron2004asymptotically, congedo2008blind}.

\paragraph{Estimating fewer sources than sensors using PCA}

Contrarily to SMICA, algorithms described above can only estimate as many sources as sensors.
Therefore, in order to estimate fewer sources, a dimension reduction step should be performed beforehand.
Principal Component Analysis is the algorithm of choice for this task.
The components are chosen to explain as much variance in the data as possible.
Since this method is blind to higher order interactions, it might discard some sources which are important but of low power.
As a consequence, \citet{artoni2018applying} argue that applying PCA before ICA leads to degraded decomposition.

\paragraph{Numerical setup}
In our experiments, we use Infomax as the reference non-Gaussian ICA algorithm with the fast and robust optimization algorithm Picard~\cite{ablin-etal:2017, ablin2018faster} and $\tanh$ as non-linearity.
The joint-diagonalization algorithm for SOBI is a combination of~\cite{ziehe2004fast} with a backtracking line-search.
The joint-diagonalization algorithm for JDIAG uses the fast implementation described in~\cite{ablin2018beyond}.
In all experiments we set the frequency bins $F_b$ of SMICA and JDIAG as uniform in the range $1-70$ Hz, with $40$ bins.
The M/EEG analysis is carried using the Python package MNE~\cite{gramfort2014mne, mnepython}.
Figures are made using Matplotlib~\cite{hunter2007matplotlib}.

The python code for SMICA is available online at \url{https://github.com/pierreablin/smica}.
\subsection{Qualitative comparisons}

\subsubsection{Comparisons on a MEG dataset}
\label{sec:expe_meg}
We start by showing the decomposition found by SMICA, JDIAG, SOBI and Infomax on a MEG dataset, where the subject was presented checkerboard patterns into the left and right visual fields, as well as monaural auditory tones to the left or right ears. Stimuli occurred every 750\,ms (See \cite{gramfort2014mne} for a description of the dataset).
MEG is acquired with 102 magnetometers and 204 gradiometers.

\begin{figure}
    \centering
    \begin{subfigure}[b]{0.23\textwidth}
        \includegraphics[width=\textwidth]{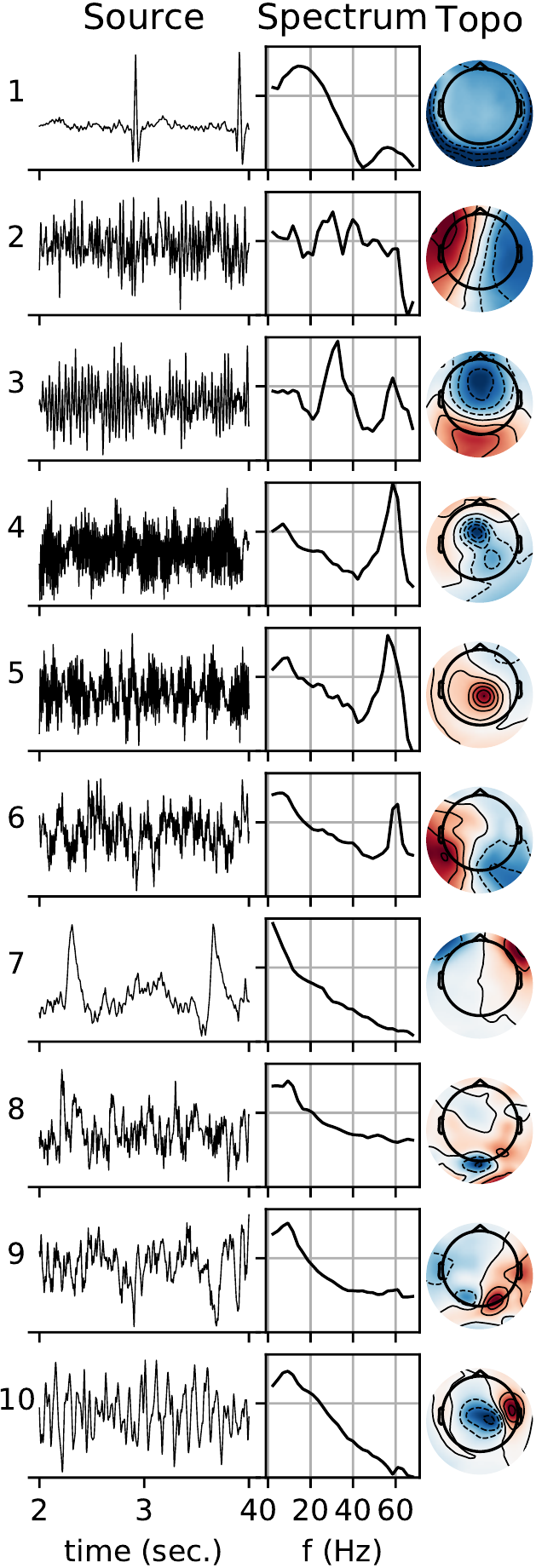}
        \caption{SMICA}

    \end{subfigure}
    ~
    \begin{subfigure}[b]{0.23\textwidth}
        \includegraphics[width=\textwidth]{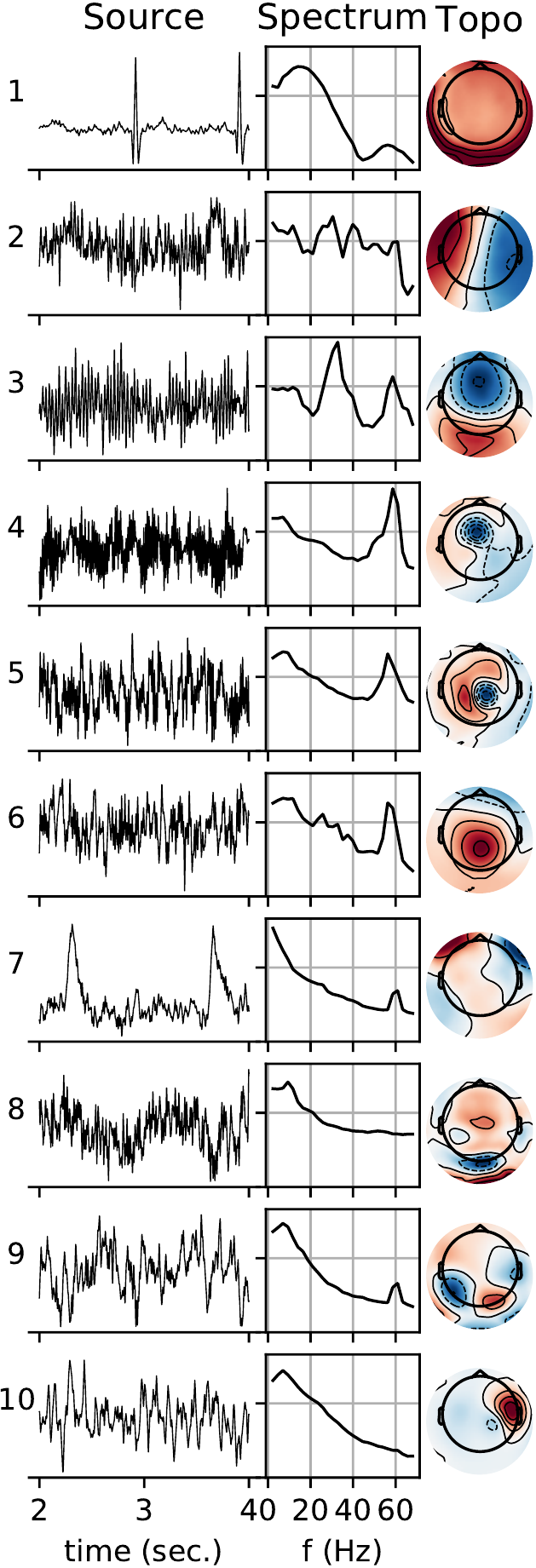}
        \caption{JDIAG}

    \end{subfigure}
    ~
    \begin{subfigure}[b]{0.23\textwidth}
        \includegraphics[width=\textwidth]{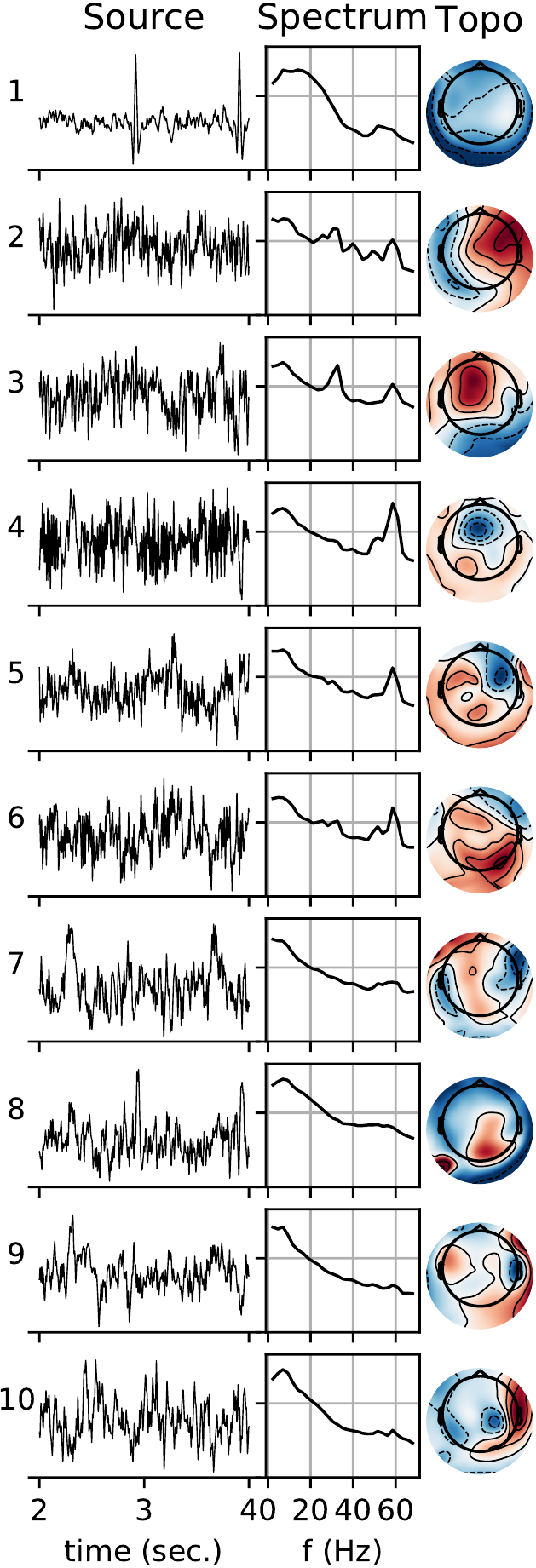}
        \caption{SOBI}

    \end{subfigure}
    ~
    \begin{subfigure}[b]{0.23\textwidth}
        \includegraphics[width=\textwidth]{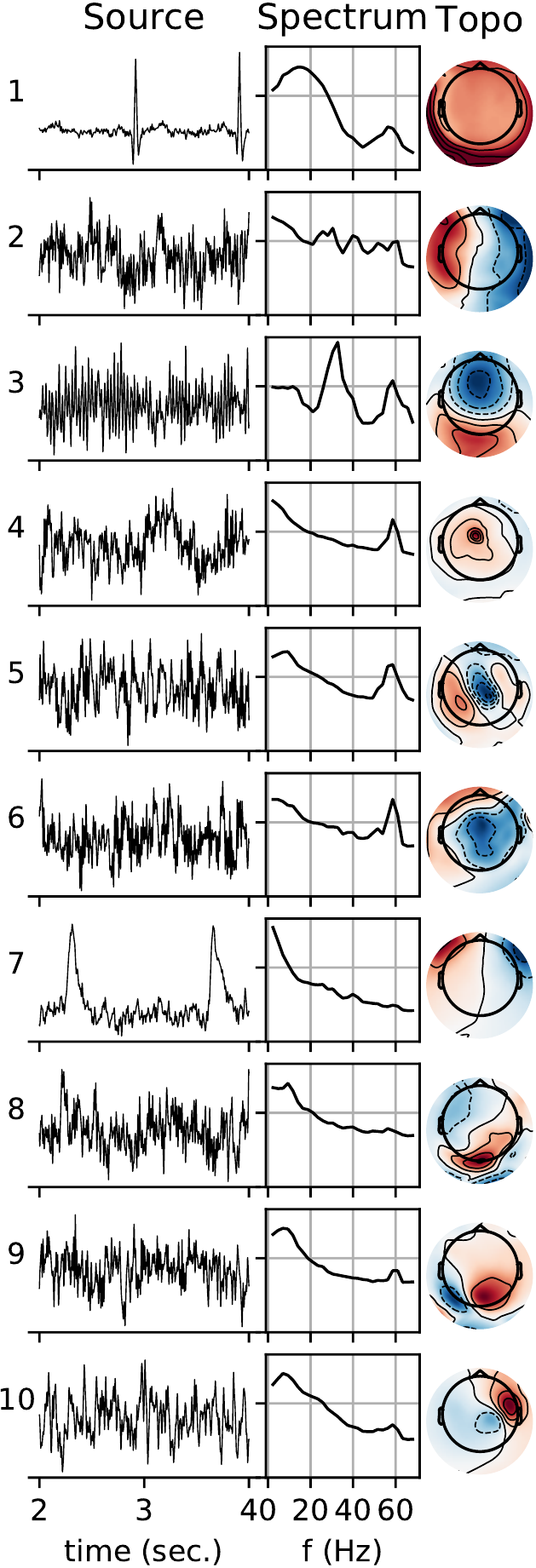}
        \caption{Infomax}

    \end{subfigure}
    \caption{Different ICA decompositions on MEG data. Source 1 corresponds to heartbeats, sources 2-6 to environmental noise, with strong peaks around 60\,Hz. Source 7 corresponds to eye blinks. Sources 8, 9 correspond to occipital alpha rhythm (excect for SOBI which did not find such sources). Source 10 corresponds to a dipolar auditory source. Sources 7, 9 found by JDIAG present artifactual 60\,Hz components, like source 1, 9 for Infomax. SOBI does not properly identify the eye-blink source.}\label{fig:ica_meg}
\end{figure}

For this experiment, we only consider the 102 magnetometer channels. Each ICA algorithm returns 40 sources (after PCA for JDIAG, SOBI and Infomax).
We hand pick $10$ sources to display in Figure~\ref{fig:ica_meg}, which shows their time-course, power spectrum and topography.

SMICA isolates heartbeats (source 1), eye blinks (source 7) and brain (sources 8-10) from line noise: only source 9 shows a very small peak at $60$ Hz. Source 7 and 9 in JDIAG's decomposition, source 1, 7, 9, 10 for SOBI and source 1, 8, 9 for Infomax do show a higher peak at 60 Hz suggesting that line noise leaks into these others sources.
SOBI fails to separate properly the eye-blinks, and gives decompositions that differ substantially from other algorithms.
This first experiment demonstrates that SMICA does reveal expected artifactual sources, both physiological and environmental, and that they potentially leak less into the valuable neural ones. More quantitative evidence is provided below.

\subsubsection{Comparisons on an EEG dataset}
\label{sec:expe_eeg}

\begin{figure}
    \centering
    \begin{subfigure}[b]{0.49\textwidth}
        \includegraphics[width=0.48\textwidth]{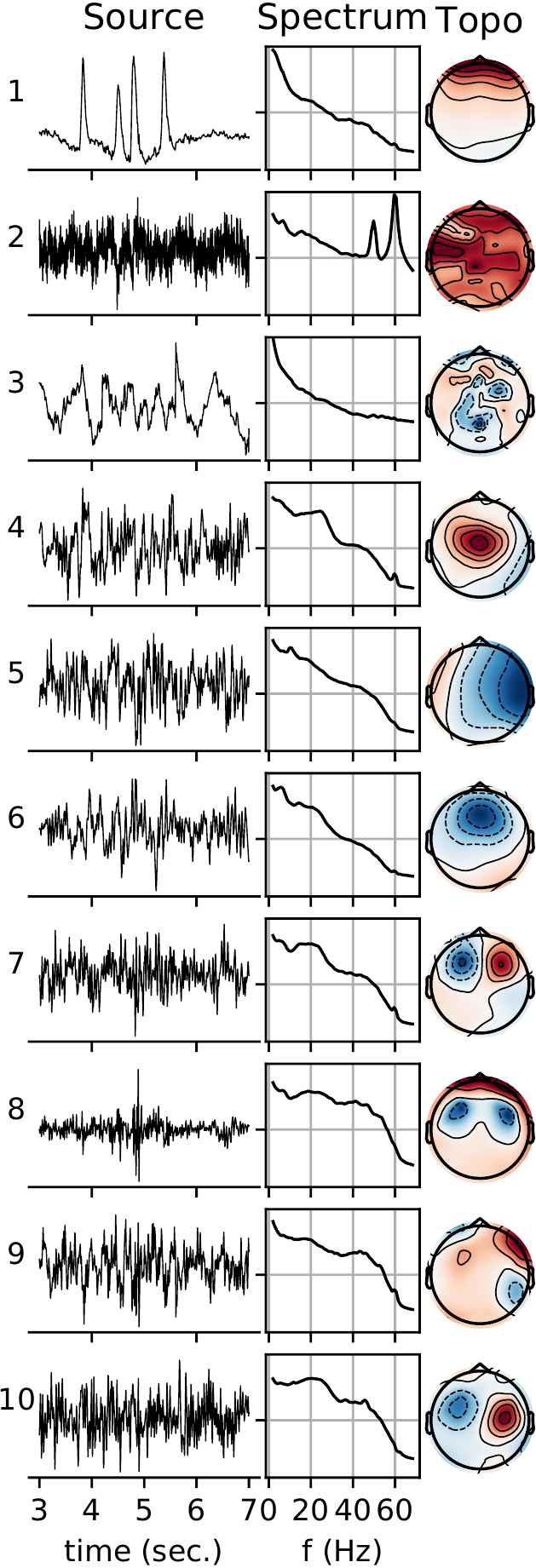}
        \includegraphics[width=0.48\textwidth]{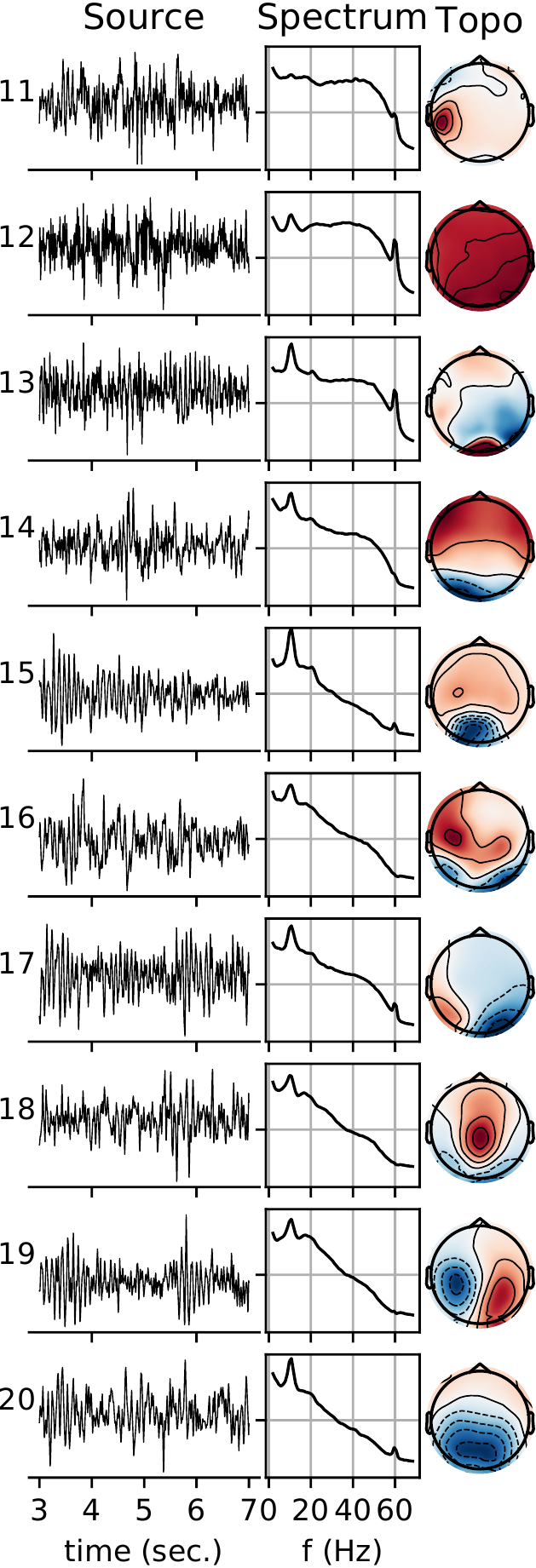}
        \caption{SMICA}
    \end{subfigure}
    \begin{subfigure}[b]{0.49\textwidth}
        \includegraphics[width=0.48\textwidth]{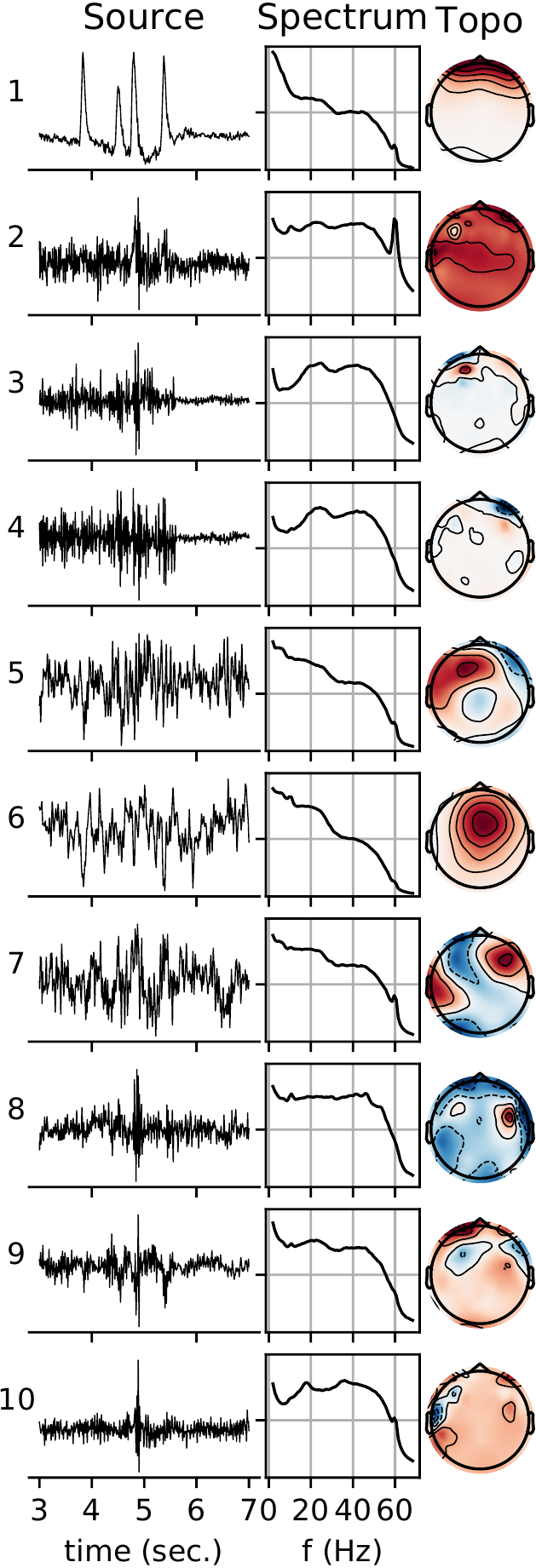}
        \includegraphics[width=0.48\textwidth]{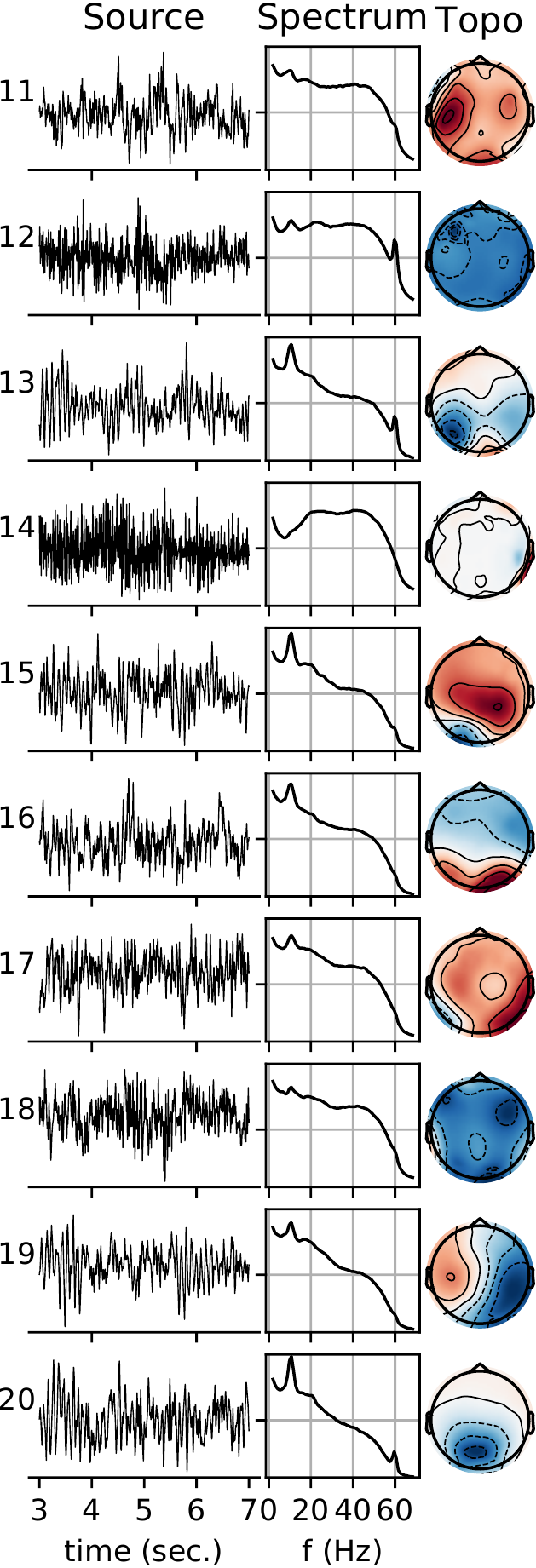}
        \caption{JDIAG}
    \end{subfigure}
\\
    \begin{subfigure}[b]{0.5\textwidth}
\includegraphics[width=.98\columnwidth]{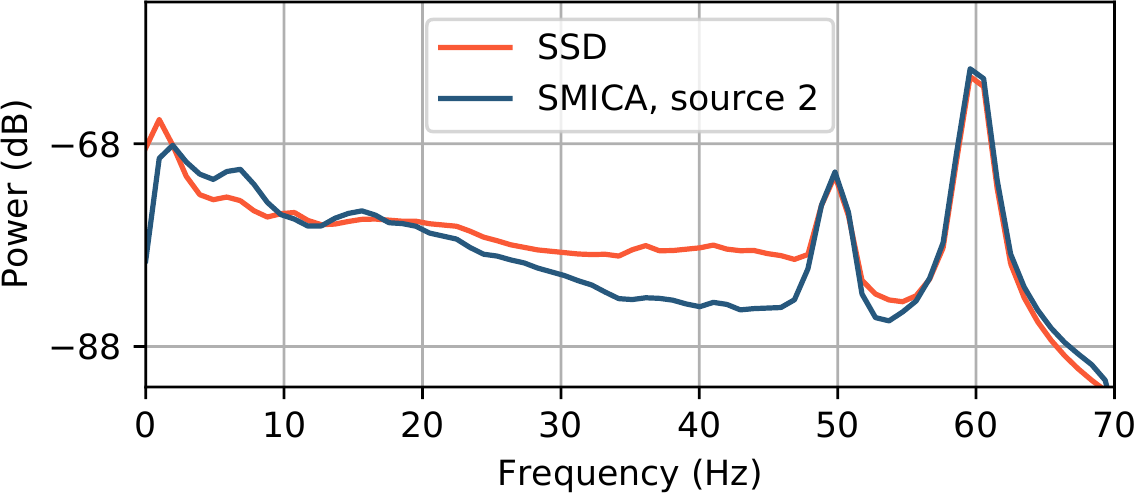}
        \caption{Spectrum of source $2$ of SMICA and source obtained by Spatio-Spectral Decomposition (SSD) at $60$Hz.}
    \end{subfigure}

    \caption{Comparison of SMICA and JDIAG on an EEG dataset. Both algorithms return 20 sources. The ordering of the sources of SMICA is made by hand; the ordering of the sources of JDIAG is made by maximizing the correlation with SMICA's sources. Both algorithms accurately recover the eye blinks (source 1). For SMICA, sources 4-20 correspond to brain activity. SMICA finds two dipolar beta-rhythm sources (sources 7 and 10). Each algorithm recovers a source corresponding to line noise, with a large peak at 60 Hz (source 2). For SMICA, there is an additional peak at $50$ Hz, which is not an artifact. Spatio-Spectral Decomposition (SSD) tuned to the source of maximal power at $60$ Hz yields a similar peak at $50$ Hz (c). }\label{fig:ica_eeg}
\end{figure}

We run SMICA and JDIAG on a 69-channel EEG data coming from the dataset described in~\cite{delorme2012independent}. Both algorithms return 20 sources, which are displayed in Figure~\ref{fig:ica_eeg}.
Differences between SMICA and JDIAG are now more striking, probably due to the greater noise level compared to the MEG recording.

Although decompositions differ in many aspects, we want to focus on source $2$ recovered by SMICA, which is not found by JDIAG. There is a sharp peak at $60$ Hz, indicating that it likely corresponds to line noise.  The second peak at $50$ Hz may seem spurious but is most probably due to spectral aliasing.  Indeed the fifth harmonic of a line at 60 Hz when sampled at 250 Hz appears at frequency $5*60-250 = 50\,\text{Hz}$.
To test whether it is a plausible line source, we resort to a separate experiment: we determine the spatial filter $w\in \bbR^{59}$ such that the time series $wX$ is of power $1$ with a maximal power at $60$ Hz. This method is called Spatio-Spectral Decomposition (SSD)~\citep{nikulin2011novel}. To do so, we filter $X$ in a narrow band around $60$ Hz, yielding signals $X_f$. Then,  we find $w$ by maximizing the power of $wX_f$ under the constraint that the power $wX$ is $1$ which is done by maximizing the Rayleigh quotient between the covariance of $X_f$ and of $X$.

The power spectrum of the corresponding source $wX$ is displayed in Figure~\ref{fig:ica_eeg}, along with the power spectrum of the source number $2$ found by SMICA.
The $50$ Hz aliased harmonic is also recovered by SSD, suggesting that the source recovered by SMICA isolates the line signal correctly.

\subsection{Quantitative results on large datasets}

\subsubsection{Experiment on MEG Phantom data}

In this section, we experiment with MEG Phantom data.
The recording comes from a fake plastic head with electric dipoles.
Dipoles emit sinusoidal pulses at a fixed frequency $20$ Hz for $0.5$ second ($10$ periods) and are then turned off for $1$ second.
This is repeated for $150$ seconds.

We have $24$ datasets. Each dataset corresponds to one dipole location among $8$ and one amplitude (either $1000$, $200$ or $20$\,nAm). The amplitude corresponds to the peak-to-peak difference.
We cut each dataset in half to obtain twice as many datasets.

As the true locations of the sources in the phantom are known one can map the origin of an ICA source by fitting an equivalent current dipole (ECD) to the source topographies and evaluate the localization errors.
On each dataset, we apply SMICA, Infomax and JDIAG to obtain $40$ sources.
For each source, we fit an ECD, and only keep the source corresponding to the closest location to the true dipole.

Besides ICA, Maxwell filtering can also be used for dipole localization. After Maxwell filtering, we compute the evoked potential and fit a dipole at the peak.

Finally, we employ Spatio-Spectral Decomposition (SSD), as described in section~\ref{sec:expe}, to find a linear combination of sensors with maximal amplitude at $20$ Hz.
This method incorporates more knowledge of the problem than others, because we provide it with the dipole frequency.

\begin{figure}
    \centering
    \includegraphics[width=\columnwidth]{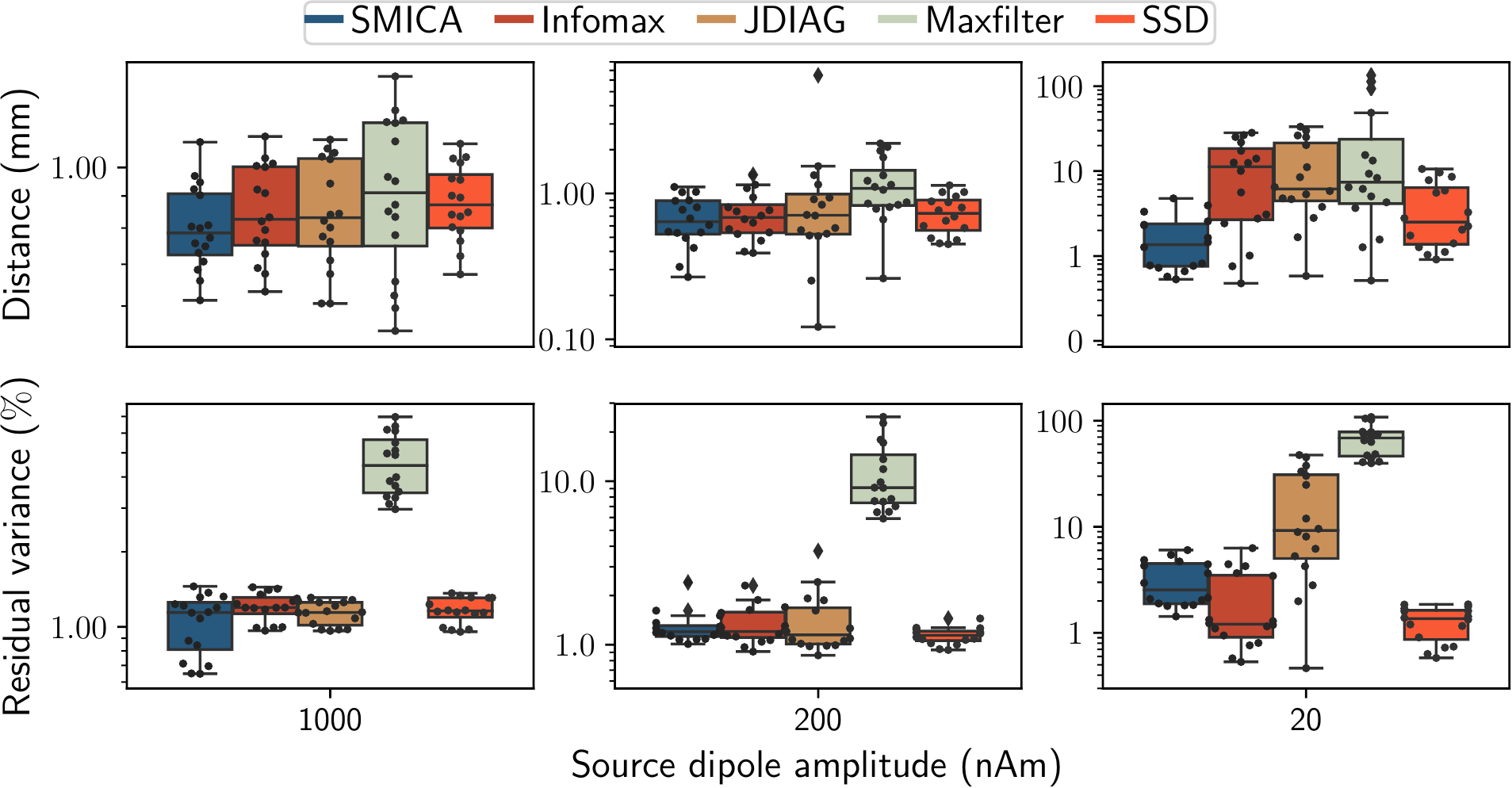}
    \caption{Dipole localization on MEG Phantom dataset. \textbf{Top}: Distance between the true dipole and the estimated dipole. \textbf{Bottom}: Residual variance in the dipole fit. Each column corresponds to a different source amplitude (from left to right, $1000$, $200$ and $20$\,nAm). Dipole fitting is applied on $24$ datasets. Each black dot corresponds to a dataset.
    When the source dipole amplitude is low ($20$\,nAm), SMICA has better localization performance than the other methods.
    }
    \label{fig:phantom}
\end{figure}

In Figure~\ref{fig:phantom}, we display the average distance to the true dipole and the residual variance in the dipole fit, for each dipole amplitude ($1000$, $200$ and $20$\,nAm).
The localization errors increase as the sources amplitude diminishes for each algorithm.
Yet, for the smallest amplitude ($20$ nAm), which is the most challenging, SMICA outperforms all other methods in terms of localization (note the logarithmic scale).

\subsection{SMICA finds highly dipolar and independent source subspaces}
\label{sec:expe_dip}

In this section, we illustrate the ability of SMICA to capture meaningful brain sources.
We use the same datasets as in~\cite{delorme2012independent}.
It contains the EEG recording of 15 subjects, with 69 EEG channels.
For a target number of independent sources, different ICA procedures described in the article are applied to the datasets.
The methods Wiener and Pseudo-Inverse correspond to the combination of SMICA with Infomax, as described in the paragraph `Combining SMICA with another ICA algorithm', section~\ref{sec:smica_in_depth}. Wiener corresponds to computing SMICA's sources with Wiener filtering, Pseudo-Inverse corresponds to computing SMICA's sources with pseudo-inversion of the mixing matrix $A$, as described in the paragraph `Source estimation by Wiener filtering', Sec.~\ref{sec:smica_in_depth}.
For each decomposition, we compute the dipolarity of each source, as well as the pairwise mutual-information between each pair of sources.

\begin{figure}
    \centering
    \includegraphics[width=\columnwidth]{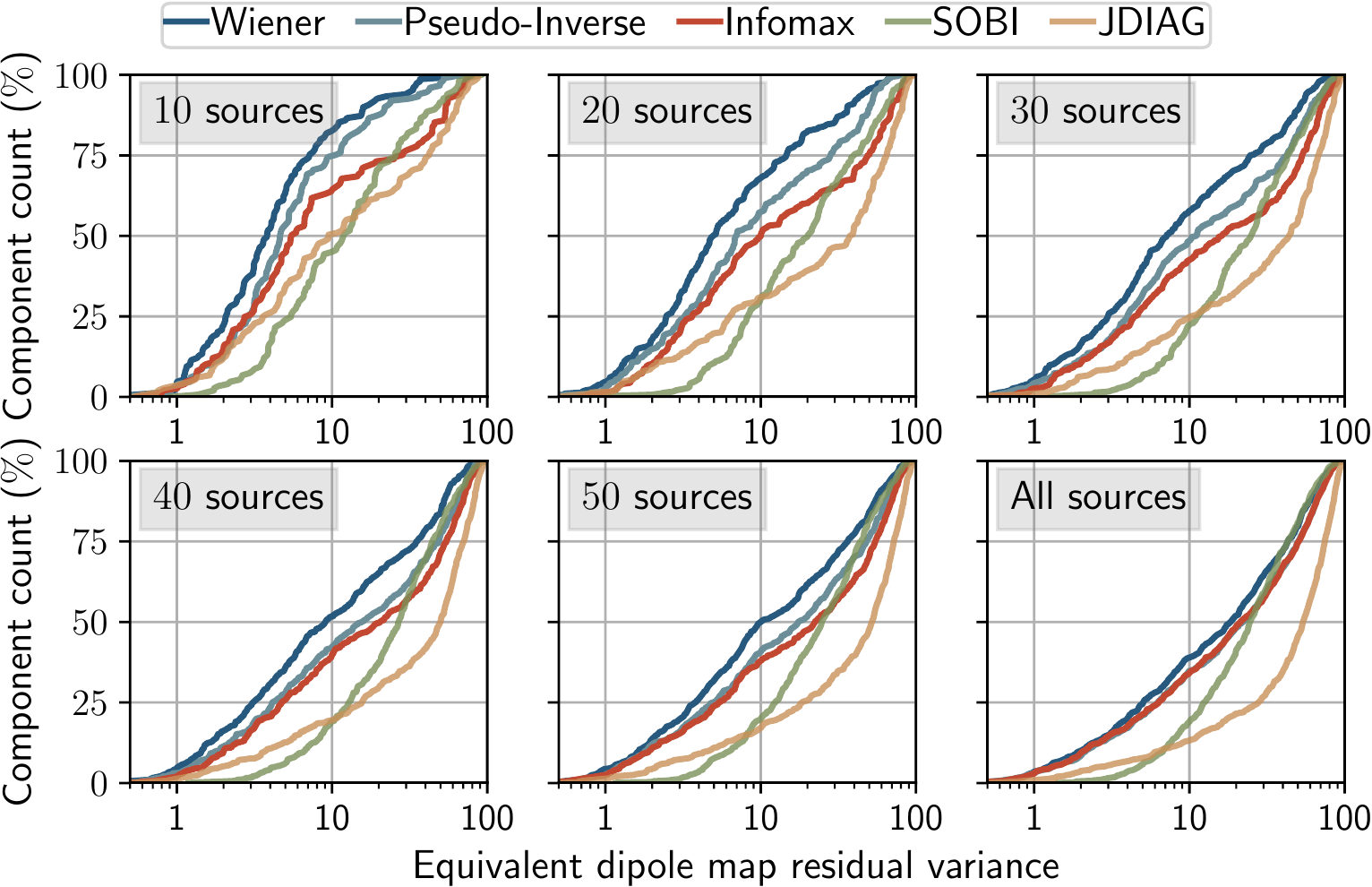}
    \caption{Distribution of equivalent dipole map residual variance for each source returned by each algorithm on the 15 datasets of 69 sensors, where each algorithm returns 10, 20, 30, 40, 50 or 69 sources.
    For Infomax, SOBI and JDIAG, PCA is first applied on the data matrix to obtain the desired number of channels.
    Wiener and Pseudo-Inverse correspond to applying Infomax on the sources recovered by SMICA, either by the Wiener or Pseudo-Inverse method.
    The figure should be understood in the following way.
    Looking at the first plot, corresponding to 10 sources, we see that about $80 \%$ of components found by the method `Wiener' have an equivalent dipole map residual variance lower than $10 \%$.
    About $70\%$ of components found by `Infomax' have an equivalent dipole map residual variance lower than $10\%$. Overall, the method `Wiener' finds more dipolar components, for every number of sources considered.
    }
    \label{fig:dip}
\end{figure}

\begin{figure}
    \centering
    \includegraphics[width=\columnwidth]{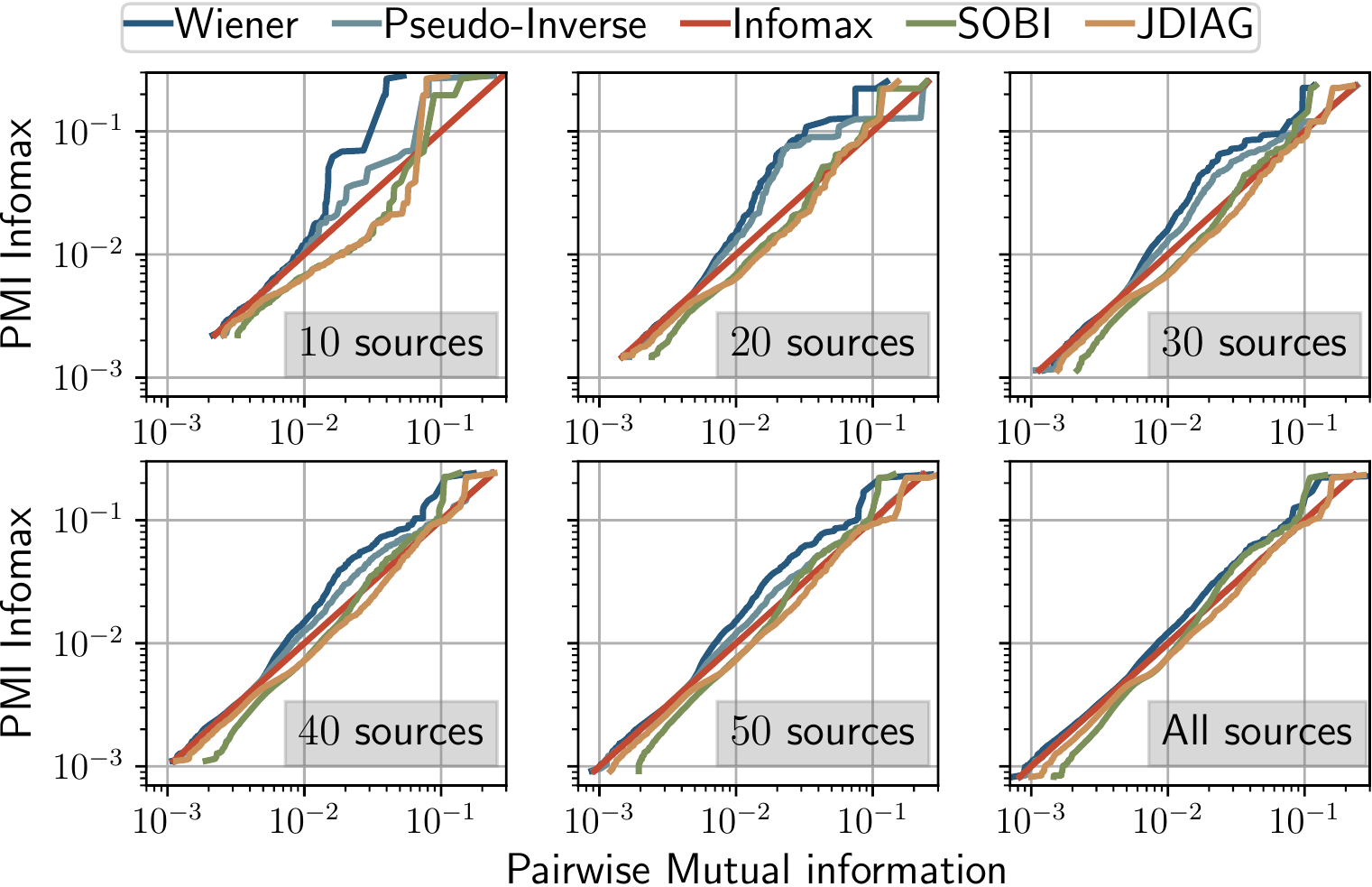}
    \caption{Pairwise mutual information (PMI) for different ICA decompositions returned by each algorithm on the 15 datasets of 69 sensors, where each algorithm returns 10, 20, 30, 40, 50 or 69 sources.
    For Infomax, SOBI and JDIAG, PCA is first applied on the data matrix to obtain the desired number of channels.
    Wiener and Pseudo-Inverse correspond to applying Infomax on the sources recovered by SMICA, either by the Wiener or Pseudo-Inverse method.
    PMI is displayed in a Q-Q plot showing the quantiles of the distribution of PMI found by each algorithm against the quantiles of distribution of PMI found by Infomax. Infomax therefore corresponds to the line $x = y$. Algorithms above the line $x=y$ have less PMI remaining than Infomax, and hence obtain more independent sources than Infomax as quantified by PMI.
    Here again the the Wiener approach is particularly competitive, especially when few sources are estimated.
    }
    \label{fig:pmi}
\end{figure}

Results for the dipolarity are displayed in Figure~\ref{fig:dip}, results for the pairwise mutual information are displayed in Figure~\ref{fig:pmi}.

The results of raw SMICA are not displayed here, as it recovered less dipolar sources than Infomax.  That is not unexpected if one follows the discussion of ~\citet{delorme2012independent}, arguing that maximum-likelihood non-Gaussian methods, like Infomax, are the ICA methods that recover the most dipolar sources.

For this experiment, using SMICA as a subspace identifier to perform dimension reduction yields the best results, regardless of the number of sources that are recovered.
The benefits of the Wiener filtering over pseudo-inversion are also illustrated: even without dimension reduction, it denoises the signals, which leads to improved ICA decompositions.

\section{Discussion}
Non-Gaussian ICA is routinely applied by many practitioners on EEG, MEG and even fMRI~\cite{beckmann-etal:05,VAROQUAUX2010288}.
In this article, we argued that using instead \emph{spectral diversity} can offer an interesting alternative.
Specifically, we proposed to use the SMICA algorithm, which leverages spectral diversity in a statistically sound manner. SMICA benefits from a noise model enabling accurate source estimation (through Wiener filtering), and the estimation of a smaller number of sources than sensors, without resorting to ad-hoc dimension reduction methods such as principal component analysis.

Results in Sec.~\ref{sec:expe_meg}~(Figure~\ref{fig:ica_meg}) have shown that SMICA can better isolate stationary environmental noise than non-Gaussian ICA methods. This can be explained by the fact that noise and brain sources usually have very different spectra. Noise tends to have peaks or bumps in the spectra in higher frequencies while neural sources have spectra with power laws or exponential decays, so called $1/f$ spectral densities~\cite{Buzsaki1926,Dehghani2010,LAROCCA2018175}.
As a consequence, this method is well suited as noise suppressor. %
It also clearly reveals physiological sources with $1/f$ power spectrum, making it easier to identify components which are artifacts.

Thanks to the noise model, we have a principled way to perform dimension reduction and to recover the source time courses. In \cite{artoni2018applying}, it is argued that PCA is suboptimal for EEG data, and that even channel subsampling is to be preferred.
In Sec.~\ref{sec:expe_dip} (Figure~\ref{fig:dip} and~\ref{fig:pmi}) we show that SMICA identifies a source subspace that contains more dipolar and independent components than PCA. As such, SMICA can be a useful tool for dimension reduction.

Importantly, as illustrated by Figure~\ref{fig:ica_meg}, SMICA is similar to JDIAG for clean data, since the noise subspace in this case is simple to find. Therefore, SMICA is more likely to be useful for  noisy datasets such as clinical recordings.

This work is not the first using and highlighting the benefits of spectral ICA for M/EEG recordings.
Some works have focused on convolutional mixtures (for which using the Fourier domain turns convolutions into products). In \cite{dyrholm-etal:07}, authors use convolutive ICA in time-domain, while \cite{ANEMULLER20031311} use a complex Infomax to find travelling waves. In the end, these methods estimate one mixing matrix for each frequency bin. However, independent models in each band might fail to recover brain rhythms with several frequency peaks. For instance, mu-rhythm is characterized by concurrent activity near 10 Hz and 20 Hz~\cite{HARI199744,Niedermeyer-da-Silva:05}.

Some improvements to the current SMICA algorithm can be investigated. First, since it comes from a principled statistical framework, it would be interesting to implement a data-driven way of computing the most likely number of sources in the data: an algorithm to automatically select the correct number of sources. However, preliminary experiments show that usual statistical criteria like Akaike Information Criterion or Bayesian Information Criterion are not satisfactory in this setting, likely because the model is not complex enough to explain fully M/EEG signals.
The EM algorithm for fitting SMICA is also quite slow, some improvements could be possible by further studying the geometry of the cost function and proposing quasi-Newton algorithms, as done recently for Infomax~\cite{ablin-etal:2017}.

\paragraph{Possible extensions of SMICA}

SMICA could be extended in several interesting way. In MEG acquisition, the empty room is sometimes recorded before the experiment. In this case, we could assume that the noise term in Eq.~\eqref{eq:ica_plus_noise} shares the same spectral signature as the empty room recording: the matrices $\Sigma_b$ are no longer estimated by the model, but are now taken as the spectral covariances of the empty room. Only the mixing matrix $A$ and the source powers $P_b$ are left to estimate.
The sources estimated by the algorithm should then only correspond to biological sources; in particular they should automatically be cleaned from line noise.

In order to obtain a pure subspace identification method, we could also drop the independence hypothesis of the sources, and no longer assume that the spectral covariance of the sources $P_b$ are diagonal.
In this setting, we can assume that $A$ is orthogonal ($A^{\top}A=I_q$). The algorithm then estimates the source subspace as $\text{Span}(A^{\top})$.

Finally, the proposed method could also be extended to non-stationary signals, where the spectral covariance matrices are replaced by time-lagged covariance matrices.
The resulting algorithm would resemble the proposed algorithm in~\cite{parra2000convolutive}, but with a proper maximum-likelihood estimation rather than the ad-hoc criterion proposed by the authors.
More generally, the spectral model with noise and EM algorithm can be employed to recover the parameters of a noisy ICA model $X= AS+N$ using any kind of second-order statistics.

\section*{Acknowledgements}

This project has received funding from the European Research Council (ERC) under the European Union’s Horizon 2020
research and innovation program (Grant agreement SLAB ERC-StG-676943).
P.A. was supported by the European Research Council (ERC project NORIA).

\section*{Competing interests statement}

This project has received funding from the European Research Council (ERC) under the European Union’s Horizon 2020
research and innovation program (Grant agreement SLAB ERC-StG-676943 and ERC project NORIA).
\section*{CRediT}

\textbf{Pierre Ablin}: Conceptualization, methodology, software, investigation, writing - original draft, visualization

\textbf{Jean-François Cardoso}: Conceptualization, methodology, software, investigation, writing - review and editing, supervision

\textbf{Alexandre Gramfort}: Conceptualization, methodology, software, writing - review and editing, supervision, funding acquisition

\appendix

\section{Spectral mismatch and likelihood}\label{sec:SMLL}

We show the statistical origin of the measure~(\ref{eq:gaussian_likelihood}) of spectral
mismatch.
The starting point is that, for a zero-mean univariate stationary times series with power
spectrum $C(f)$, the Fourier coefficients $\tilde{\bx}_k$ defined at Eq.(\ref{eq:fourier})
have zero mean and covariance $\bbE \tilde{\bx}_k\tilde{\bx}_k^H = C(k/T)$.
Next, asymptotically (for large $T$), these coefficients also are normally distributed and
pairwise uncorrelated~\cite{brillinger2001time}.
The \emph{Whittle approximation} to the likelihood consists in assuming that these
properties hold even in finite sample size.
In that case, the log density for the Fourier coefficients is just the sum of logarithmic
Gaussian densities over frequencies;
\begin{equation}\label{eq:logpFou}
  \log p(\tilde{\bx}_1, \ldots, \tilde{\bx}_{T/2})
  =
  - \sum_{k=1}^{T/2} \bigl( \Tr \bigl( C(\frac k T)\inv (x_kx_k^H) \bigr) +    \log \det(2\pi C(\frac k T)) \bigr)
\end{equation}
where the form of the summands in Eq.~(\ref{eq:logpFou}) is derived in
App.~\ref{sec:likeCvec} and is valid only because $C(f)$ is real-valued.
The next step is to approximate the spectra as constant over the spectral bands.
If $C(f)=C_b$ when $f\in I_b$, then Eq.~(\ref{eq:logpFou}) becomes
\begin{equation}
  \label{eq:whittleBand}
  \log p(\tilde{\bx}_1, \ldots, \tilde{\bx}_{T/2})
  =
  - \sum_{b=1}^B  n_b \left(  \Tr (\widehat C_b C_b\inv ) + \log\det (2\pi  C_b) \right)
  .
\end{equation}
Hence, with the definition (\ref{eq:kl}) of the KL divergence, we obtain
\begin{equation}
  \label{eq:whittleKL}
  \log p(\tilde{\bx}_1, \ldots, \tilde{\bx}_{T/2})
  =
  - \sum_{b=1}^B 2 n_b \KL\left(\hat{C}_b, C_b(\theta)\right) + \mathrm{cst}
\end{equation}
It shows that, up to a constant term, the spectral matching criterion is minus the
logarithm of the Whittle likelihood.

The presence of the factor $2$ in~(\ref{eq:whittleBand}) can be traced back to the fact
that a Fourier component at frequency $k$ has 2 degrees of freedom: its real and its
imaginary parts.

\section{Likelihood and complex vectors}\label{sec:likeCvec}

We give joint pdf for the real and imaginary parts of a zero-mean complex random $p\times 1$ vector $x=a+ib$ with a \emph{real} covariance
matrix:
\begin{displaymath}
  x = a + i b \qquad C = \bbE[ xx^H] \in \bbR^{p\times p} .
\end{displaymath}
and whose distribution is invariant under any phase change, that is, $x$ has the same
distribution as $e^{i\phi} x$ for any angle~$\phi$.  We look at the consequence of this
invariance on the structure of $C$ and give the joint pdf of $a$ and $b$ when they are
jointly Gaussian.   We start with:
\begin{align*}
  C = \bbE[ xx^H]
  &= \bbE[(a+ib)(a+ib)^H ]
  = \bbE[(aa^\top  + bb^\top) \,-\, i \ \bbE[ ab^\top - ba^\top ]
  \\
  \bbE[ xx^\top  ]
  &= \bbE[(a+ib)(a+ib)^\top ]
  = \bbE[(aa^\top -bb^\top) \,+\, i \ \bbE[ ab^\top + ba^\top ]
\end{align*}
If $x$ is changed into $e^{i\phi}x$, then matrix $\bbE[ xx^H]$ is unchanged but
$\bbE[ xx^\top]$ is changed into $e^{2i\phi} \bbE[ xx^\top]$.  However, by the phase
invariance, there should be no change.  That is only possible if $\bbE[ xx^\top]=0$.
Hence, we have $\bbE[(aa^\top -bb^\top) = 0$ and $\bbE[ ab^\top + ba^\top ]=0$.
Combining that with the assumption of a real covariance matrix $C = \bbE[ xx^H]$ which
implies $\bbE[ ab^\top - ba^\top ]=0$ yields
\begin{displaymath}
  \bbE[aa^\top] = \bbE[bb^\top] = C/2 \qquad \bbE[ ab^\top] = \bbE[ ba^\top ] = 0
\end{displaymath}
Therefore the joint $(2p)\times(2p)$ covariance matrix for the pair $(a,b)$ is
\begin{displaymath}
  \Cov\Bigl( \bivec a b\Bigr) =
  \bimat
  {\bbE[aa^\top]}
  {\bbE[ab^\top]}
  {\bbE[ba^\top]}
  {\bbE[bb^\top]}
  =
  \bimat{C/2} 0 0 {C/2}
\end{displaymath}
If $a$ and $b$ are jointly Gaussian, their probability density $p(a,b)$ is given by
\begin{displaymath}
  -2\log p(a,b) =  \bivec a b ^\top \Sigma\inv  \bivec a b  + \log \det(2\pi\Sigma)
  \qquad
  \Sigma = \Cov\Bigl( \bivec a b\Bigr) =   \bimat{C/2} 0 0 {C/2}  .
\end{displaymath}
The block structure of $\Sigma$ yields $ \log \det(2\pi\Sigma) = 2 \log \det(\pi C) $ and also
\begin{displaymath}
  \bivec a b ^\top   \bimat{C/2} 0 0 {C/2}  \inv  \bivec a b
  =
  2 \Tr \bigl( C\inv (aa^\top+bb^\top) \bigr)
  =
  2 \Tr \bigl( C\inv \Real(xx^H) \bigr)
  .
\end{displaymath}
The desired result thus is
\begin{displaymath}
  -2\log p(a,b) = 2 \bigl( \Tr \bigl( C\inv \Real(xx^H) \bigr) +    \log \det(C) \bigr) \,+\, 2p\log\pi .
\end{displaymath}

\section{The EM algorithm for SMICA}\label{sec:em_smica}

The parameters are $\theta = \{A, \Sigma_1, \dots, \Sigma_B, P_1, \dots, P_B\}$, and the latent variables are the sources in each frequency bands $S_1, \dots, S_B$.

\paragraph{E-step}

At the E-step, the sufficient statistics of the model are computed. Since the model is Gaussian, they are simply the second-order statistics: $\bbE[S_bS_b^{\top}|\theta], \bbE[S_bX_b^{\top} |\theta]$ and $\bbE[X_bX_b^{\top}|\theta]$.
In the following, let $\Gamma_b = (A^{\top} \Sigma_b^{-1}A + P_b^{-1})^{-1}$ and $W_b = \Gamma_bA^{\top}\Sigma_b^{-1}$ the Wiener filter. We have:
\begin{equation}
    R^{XX}_b \triangleq \bbE[X_bX_b^{\top}|\theta] = \hat{C}_b
    \label{eq:cov_xx}
\end{equation}
\begin{equation}
    R^{SX}_b \triangleq\bbE[S_bX_b^{\top}|\theta] = W_b \hat{C}_b
    \label{eq:cov_sx}
\end{equation}
\begin{equation}
    R^{SS}_b \triangleq\bbE[S_bS_b^{\top}|\theta] = W_b \hat{C}_bW_b^{\top} + \Gamma_b
    \label{eq:cov_ss}
\end{equation}

\paragraph{M-step}

At the M-step, the parameters of the model $\theta$ should be modified in order to decrease the loss function, using the sufficient statistics obtained in the E-step.
The EM functional writes:
\begin{equation}
    \Phi  = \sum_{b=1}^B n_b [\Tr((R^{XX}_b - 2A R^{SX}_b + A R^{SS}_bA^{\top})\Sigma_b^{-1}) + \Tr(R^{SS}_bP_b^{-1}) + \log|\Sigma_b| + \log|P_b| ] \enspace,
    \label{eq:em_fun}
\end{equation}
which should be minimized with respect to the parameters $\theta$.

\begin{itemize}
    \item Optimizing $P_b$: the source powers are decoupled from the other parameters in~\eqref{eq:em_fun}. Minimization of $\Phi$ w.r.t. $P_b$ is easily obtained by canceling the gradient, yielding: $$P_b = \text{diag}(R^{SS}_b) \enspace .$$
    \item Optimizing $\Sigma_b$: the mixing matrix $A$ and the noise covariance are entangled in eq.~\eqref{eq:em_fun}, rendering the analytic minimization of $\Phi$ impossible. Therefore, we first minimize $\Phi$ w.r.t $\Sigma_b$, keeping $A$ constant.
    This yields:
    $$
        \Sigma_b = \text{diag}(R^{XX}_b - 2AR^{SX}_b +AR^{SS}_bA^{\top}) \enspace .
    $$
    \item Optimizing $A$: keeping the noise levels fixed, minimizing $\Phi$ w.r.t. $A$ yields, by canceling the gradient: $ \sum_{i=1}^B n_b\Sigma_b^{-1}(R^{XS}_b - AR^{SS}_b)=0$.
    This can be seen as a system of equations for the rows of $A$ which, thanks to the diagonality of $\Sigma_b$, is easily seen to decouple across the rows.   For each row, simple algebra yields the close form solution:
     $$
    \text{for} \enspace r=1\dots p,
    \enspace A_{r:} = Q_{r:} M_r^{-1}
    \quad\text{with}
    \quad Q = \sum_{b=1}^Bn_b \Sigma_b^{-1} R^{XS}_b
    \quad M_r = \sum_{b=1}^B n_b \sigma_{i, r}^{-2} R_b^{SS}.
    $$
    Therefore, the EM update of $A$ only requires solving $p$ linear systems of size $q\times q$.
\end{itemize}

\paragraph{Implementation details}

The EM algorithm iterates the E and M step until a certain convergence criterion is reached.
In practice, iterations are stopped when the difference between two consecutive values of the log-likelihood is below a threshold: $\mathcal{L}^{t+1} > \mathcal{L}^t -  \varepsilon$.
In order to have a good initialization for the algorithm, we first fit the model with a fixed noise level for each bin: we estimate $ \Sigma$  subject to $\Sigma_b = \Sigma$ for all $i$. In this setting, the M-step is much simpler and computationally quicker. Then, the core SMICA algorithm with free noise starts with $\Sigma_b$ all equal to the estimated noise level, and $A$
and $P_b$ start from the same initial value.

\section*{References}
\bibliographystyle{apa}
\bibliography{references}

\begin{thebibliography}{}

\bibitem[\protect\astroncite{Ablin et~al.}{2018a}]{ablin2018faster}
Ablin, P., Cardoso, J.-F., and Gramfort, A. (2018a).
\newblock Faster {ICA} under orthogonal constraint.
\newblock In {\em 2018 IEEE International Conference on Acoustics, Speech and
  Signal Processing (ICASSP)}, pages 4464--4468. IEEE.

\bibitem[\protect\astroncite{Ablin et~al.}{2018b}]{ablin-etal:2017}
Ablin, P., Cardoso, J.-F., and Gramfort, A. (2018b).
\newblock Faster independent component analysis by preconditioning with
  {H}essian approximations.
\newblock {\em IEEE Transactions on Signal Processing}, 66(15):4040--4049.

\bibitem[\protect\astroncite{Ablin et~al.}{2019}]{ablin2018beyond}
Ablin, P., Cardoso, J.-F., and Gramfort, A. (2019).
\newblock {Beyond Pham's algorithm for joint diagonalization}.
\newblock In {\em European Symposium on Artificial Neural Networks,
  Computational Intelligence and Machine Learning (ESANN)}.

\bibitem[\protect\astroncite{Anemüller et~al.}{2003}]{ANEMULLER20031311}
Anemüller, J., Sejnowski, T.~J., and Makeig, S. (2003).
\newblock Complex independent component analysis of frequency-domain
  electroencephalographic data.
\newblock {\em Neural Networks}, 16(9):1311 -- 1323.
\newblock Neuroinformatics.

\bibitem[\protect\astroncite{Artoni et~al.}{2018}]{artoni2018applying}
Artoni, F., Delorme, A., and Makeig, S. (2018).
\newblock Applying dimension reduction to {EEG} data by principal component
  analysis reduces the quality of its subsequent independent component
  decomposition.
\newblock {\em NeuroImage}, 175:176--187.

\bibitem[\protect\astroncite{Barth{\'e}lemy
  et~al.}{2013}]{barthelemy2013multivariate}
Barth{\'e}lemy, Q., Gouy-Pailler, C., Isaac, Y., Souloumiac, A., Larue, A., and
  Mars, J.~I. (2013).
\newblock Multivariate temporal dictionary learning for {EEG}.
\newblock {\em Journal of neuroscience methods}, 215(1):19--28.

\bibitem[\protect\astroncite{Beckmann et~al.}{2005}]{beckmann-etal:05}
Beckmann, C.~F., DeLuca, M., Devlin, J.~T., and Smith, S.~M. (2005).
\newblock Investigations into resting-state connectivity using independent
  component analysis.
\newblock {\em Philosophical Transactions of the Royal Society B: Biological
  Sciences}, 360(1457):1001--1013.

\bibitem[\protect\astroncite{Bell and Sejnowski}{1995}]{bell1995information}
Bell, A.~J. and Sejnowski, T.~J. (1995).
\newblock An information-maximization approach to blind separation and blind
  deconvolution.
\newblock {\em Neural computation}, 7(6):1129--1159.

\bibitem[\protect\astroncite{Belouchrani et~al.}{1997}]{belouchrani1997blind}
Belouchrani, A., Abed-Meraim, K., Cardoso, J.-F., and Moulines, E. (1997).
\newblock A blind source separation technique using second-order statistics.
\newblock {\em IEEE Transactions on signal processing}, 45(2):434--444.

\bibitem[\protect\astroncite{Brillinger}{2001}]{brillinger2001time}
Brillinger, D.~R. (2001).
\newblock {\em Time series: data analysis and theory}.
\newblock SIAM.

\bibitem[\protect\astroncite{Brookes et~al.}{2011}]{brookes2011investigating}
Brookes, M.~J., Woolrich, M., Luckhoo, H., Price, D., Hale, J.~R., Stephenson,
  M.~C., Barnes, G.~R., Smith, S.~M., and Morris, P.~G. (2011).
\newblock Investigating the electrophysiological basis of resting state
  networks using magnetoencephalography.
\newblock {\em Proceedings of the National Academy of Sciences},
  108(40):16783--16788.

\bibitem[\protect\astroncite{Brown et~al.}{1992}]{brown1992introduction}
Brown, R.~G., Hwang, P.~Y., et~al. (1992).
\newblock {\em Introduction to random signals and applied Kalman filtering},
  volume~3.
\newblock Wiley New York.

\bibitem[\protect\astroncite{Buzs{\'a}ki and Draguhn}{2004}]{Buzsaki1926}
Buzs{\'a}ki, G. and Draguhn, A. (2004).
\newblock Neuronal oscillations in cortical networks.
\newblock {\em Science}, 304(5679):1926--1929.

\bibitem[\protect\astroncite{Cardoso et~al.}{2002}]{cardoso2002blind}
Cardoso, J.-F., Snoussi, H., and Delabrouille, J. (2002).
\newblock Blind separation of noisy {G}aussian stationary sources. application
  to cosmic microwave background imaging.
\newblock In {\em 2002 11th European Signal Processing Conference}, pages 1--4.
  IEEE.

\bibitem[\protect\astroncite{Cardoso and Souloumiac}{1993}]{cardoso1993blind}
Cardoso, J.-F. and Souloumiac, A. (1993).
\newblock Blind beamforming for non-gaussian signals.
\newblock {\em IEE proceedings F (radar and signal processing)},
  140(6):362--370.

\bibitem[\protect\astroncite{Comon}{1994}]{comon1994independent}
Comon, P. (1994).
\newblock Independent component analysis, a new concept?
\newblock {\em Signal processing}, 36(3):287--314.

\bibitem[\protect\astroncite{Congedo et~al.}{2008}]{congedo2008blind}
Congedo, M., Gouy-Pailler, C., and Jutten, C. (2008).
\newblock On the blind source separation of human electroencephalogram by
  approximate joint diagonalization of second order statistics.
\newblock {\em Clinical Neurophysiology}, 119(12):2677--2686.

\bibitem[\protect\astroncite{{Dammers} et~al.}{2008}]{dammers-etal:08}
{Dammers}, J., {Schiek}, M., {Boers}, F., {Silex}, C., {Zvyagintsev}, M.,
  {Pietrzyk}, U., and {Mathiak}, K. (2008).
\newblock Integration of amplitude and phase statistics for complete artifact
  removal in independent components of neuromagnetic recordings.
\newblock {\em IEEE Transactions on Biomedical Engineering}, 55(10):2353--2362.

\bibitem[\protect\astroncite{Dehghani et~al.}{2010}]{Dehghani2010}
Dehghani, N., B{\'e}dard, C., Cash, S.~S., Halgren, E., and Destexhe, A.
  (2010).
\newblock Comparative power spectral analysis of simultaneous
  elecroencephalographic and magnetoencephalographic recordings in humans
  suggests non-resistive extracellular media.
\newblock {\em Journal of Computational Neuroscience}, 29(3):405--421.

\bibitem[\protect\astroncite{Delabrouille
  et~al.}{2003}]{delabrouille2003multidetector}
Delabrouille, J., Cardoso, J.-F., and Patanchon, G. (2003).
\newblock Multidetector multicomponent spectral matching and applications for
  cosmic microwave background data analysis.
\newblock {\em Monthly Notices of the Royal Astronomical Society},
  346(4):1089--1102.

\bibitem[\protect\astroncite{Delorme et~al.}{2012}]{delorme2012independent}
Delorme, A., Palmer, J., Onton, J., Oostenveld, R., and Makeig, S. (2012).
\newblock Independent {EEG} sources are dipolar.
\newblock {\em PloS one}, 7(2):e30135.

\bibitem[\protect\astroncite{Dempster et~al.}{1977}]{dempster1977maximum}
Dempster, A.~P., Laird, N.~M., and Rubin, D.~B. (1977).
\newblock Maximum likelihood from incomplete data via the {EM} algorithm.
\newblock {\em Journal of the Royal Statistical Society: Series B
  (Methodological)}, 39(1):1--22.

\bibitem[\protect\astroncite{Doron and Yeredor}{2004}]{doron2004asymptotically}
Doron, E. and Yeredor, A. (2004).
\newblock Asymptotically optimal blind separation of parametric {Gaussian}
  sources.
\newblock In {\em International Conference on Independent Component Analysis
  and Signal Separation}, pages 390--397. Springer.

\bibitem[\protect\astroncite{Dyrholm et~al.}{2007}]{dyrholm-etal:07}
Dyrholm, M., Makeig, S., and Hansen, L.~K. (2007).
\newblock Model selection for convolutive {ICA} with an application to
  spatiotemporal analysis of {EEG}.
\newblock {\em Neural Computation}, 19(4):934--955.

\bibitem[\protect\astroncite{G{\'o}mez-Herrero
  et~al.}{2008}]{gomez2008measuring}
G{\'o}mez-Herrero, G., Atienza, M., Egiazarian, K., and Cantero, J.~L. (2008).
\newblock Measuring directional coupling between {EEG} sources.
\newblock {\em Neuroimage}, 43(3):497--508.

\bibitem[\protect\astroncite{Gramfort et~al.}{2013}]{mnepython}
Gramfort, A., Luessi, M., Larson, E., Engemann, D., Strohmeier, D., Brodbeck,
  C., Goj, R., Jas, M., Brooks, T., Parkkonen, L., and Hämäläinen, M.
  (2013).
\newblock {MEG and EEG data analysis with MNE-Python}.
\newblock {\em Frontiers in Neuroscience}, 7:267.

\bibitem[\protect\astroncite{Gramfort et~al.}{2014}]{gramfort2014mne}
Gramfort, A., Luessi, M., Larson, E., Engemann, D.~A., Strohmeier, D.,
  Brodbeck, C., Parkkonen, L., and H{\"a}m{\"a}l{\"a}inen, M.~S. (2014).
\newblock {MNE} software for processing {MEG} and {EEG} data.
\newblock {\em Neuroimage}, 86:446--460.

\bibitem[\protect\astroncite{H{\"a}m{\"a}l{\"a}inen
  et~al.}{1993}]{hamalainen1993magnetoencephalography}
H{\"a}m{\"a}l{\"a}inen, M., Hari, R., Ilmoniemi, R.~J., Knuutila, J., and
  Lounasmaa, O.~V. (1993).
\newblock Magnetoencephalography—theory, instrumentation, and applications to
  noninvasive studies of the working human brain.
\newblock {\em Reviews of modern Physics}, 65(2):413.

\bibitem[\protect\astroncite{Hari and Salmelin}{1997}]{HARI199744}
Hari, R. and Salmelin, R. (1997).
\newblock Human cortical oscillations: a neuromagnetic view through the skull.
\newblock {\em Trends in Neurosciences}, 20(1):44 -- 49.

\bibitem[\protect\astroncite{Hunter}{2007}]{hunter2007matplotlib}
Hunter, J.~D. (2007).
\newblock Matplotlib: A 2d graphics environment.
\newblock {\em Computing in science \& engineering}, 9(3):90--95.

\bibitem[\protect\astroncite{Hyv{\"a}rinen}{1998}]{hyvarinen1998independent}
Hyv{\"a}rinen, A. (1998).
\newblock Independent component analysis in the presence of {Gaussian} noise by
  maximizing joint likelihood.
\newblock {\em Neurocomputing}, 22(1-3):49--67.

\bibitem[\protect\astroncite{Hyv{\"a}rinen}{1999}]{hyvarinen1999fast}
Hyv{\"a}rinen, A. (1999).
\newblock Fast and robust fixed-point algorithms for independent component
  analysis.
\newblock {\em IEEE transactions on Neural Networks}, 10(3):626--634.

\bibitem[\protect\astroncite{Hyv{\"a}rinen
  et~al.}{2004}]{hyvarinen2004independent}
Hyv{\"a}rinen, A., Karhunen, J., and Oja, E. (2004).
\newblock {\em Independent component analysis}, volume~46.
\newblock John Wiley \& Sons.

\bibitem[\protect\astroncite{Hyv{\"a}rinen
  et~al.}{2010}]{hyvarinen2010independent}
Hyv{\"a}rinen, A., Ramkumar, P., Parkkonen, L., and Hari, R. (2010).
\newblock Independent component analysis of short-time {F}ourier transforms for
  spontaneous {EEG/MEG} analysis.
\newblock {\em NeuroImage}, 49(1):257--271.

\bibitem[\protect\astroncite{Ikeda and Toyama}{2000}]{ikeda2000independent}
Ikeda, S. and Toyama, K. (2000).
\newblock Independent component analysis for noisy data—{MEG} data analysis.
\newblock {\em Neural Networks}, 13(10):1063--1074.

\bibitem[\protect\astroncite{Jung et~al.}{2000}]{jung2000removing}
Jung, T.-P., Makeig, S., Humphries, C., Lee, T.-W., Mckeown, M.~J., Iragui, V.,
  and Sejnowski, T.~J. (2000).
\newblock Removing electroencephalographic artifacts by blind source
  separation.
\newblock {\em Psychophysiology}, 37(2):163--178.

\bibitem[\protect\astroncite{Lee et~al.}{1999}]{lee1999independent}
Lee, T.-W., Girolami, M., and Sejnowski, T.~J. (1999).
\newblock Independent component analysis using an extended infomax algorithm
  for mixed subgaussian and supergaussian sources.
\newblock {\em Neural computation}, 11(2):417--441.

\bibitem[\protect\astroncite{Liu et~al.}{2017}]{liu2017sparse}
Liu, F., Wang, S., Rosenberger, J., Su, J., and Liu, H. (2017).
\newblock A sparse dictionary learning framework to discover discriminative
  source activations in eeg brain mapping.
\newblock In {\em Thirty-First AAAI Conference on Artificial Intelligence}.

\bibitem[\protect\astroncite{Makeig et~al.}{1996}]{makeig1996independent}
Makeig, S., Bell, A.~J., Jung, T.-P., and Sejnowski, T.~J. (1996).
\newblock Independent component analysis of electroencephalographic data.
\newblock In {\em Advances in neural information processing systems}, pages
  145--151.

\bibitem[\protect\astroncite{Makeig et~al.}{2004}]{makeig2004mining}
Makeig, S., Debener, S., Onton, J., and Delorme, A. (2004).
\newblock Mining event-related brain dynamics.
\newblock {\em Trends in cognitive sciences}, 8(5):204--210.

\bibitem[\protect\astroncite{Mantini et~al.}{2008}]{mantini2008improving}
Mantini, D., Franciotti, R., Romani, G.~L., and Pizzella, V. (2008).
\newblock Improving {MEG} source localizations: an automated method for
  complete artifact removal based on independent component analysis.
\newblock {\em NeuroImage}, 40(1):160--173.

\bibitem[\protect\astroncite{Niedermeyer and Lopes~da
  Silva}{2005}]{Niedermeyer-da-Silva:05}
Niedermeyer, E. and Lopes~da Silva, F.~H. (2005).
\newblock {\em Electroencephalography : basic principles, clinical
  applications, and related fields}.
\newblock Philadelphia ; London : Lippincott Williams \& Wilkins, 5th ed
  edition.

\bibitem[\protect\astroncite{Nikulin et~al.}{2011}]{nikulin2011novel}
Nikulin, V.~V., Nolte, G., and Curio, G. (2011).
\newblock A novel method for reliable and fast extraction of neuronal {EEG/MEG}
  oscillations on the basis of spatio-spectral decomposition.
\newblock {\em NeuroImage}, 55(4):1528--1535.

\bibitem[\protect\astroncite{Olshausen and
  Field}{1996}]{olshausen1996emergence}
Olshausen, B.~A. and Field, D.~J. (1996).
\newblock Emergence of simple-cell receptive field properties by learning a
  sparse code for natural images.
\newblock {\em Nature}, 381(6583):607--609.

\bibitem[\protect\astroncite{Palmer et~al.}{2012}]{palmer2012amica}
Palmer, J.~A., Kreutz-Delgado, K., and Makeig, S. (2012).
\newblock {AMICA}: An adaptive mixture of independent component analyzers with
  shared components.
\newblock {\em Swartz Center for Computatonal Neursoscience, University of
  California San Diego, Tech. Rep}.

\bibitem[\protect\astroncite{Parra and Spence}{2000}]{parra2000convolutive}
Parra, L. and Spence, C. (2000).
\newblock Convolutive blind separation of non-stationary sources.
\newblock {\em IEEE transactions on Speech and Audio Processing},
  8(3):320--327.

\bibitem[\protect\astroncite{Pham and Cardoso}{2003}]{spie03b}
Pham, D.-T. and Cardoso, J.-F. (2003).
\newblock Source adaptive blind source separation: {G}aussian models and
  sparsity.
\newblock In {\em Wavelets: Applications in Signal and Image Processing, X,
  Proc. of SPIE}, volume 5207, San Diego.

\bibitem[\protect\astroncite{Pham and Garat}{1997}]{pham1997blind}
Pham, D.~T. and Garat, P. (1997).
\newblock Blind separation of mixture of independent sources through a
  quasi-maximum likelihood approach.
\newblock {\em IEEE transactions on Signal Processing}, 45(7):1712--1725.

\bibitem[\protect\astroncite{Rocca et~al.}{2018}]{LAROCCA2018175}
Rocca, D.~L., Zilber, N., Abry, P., van Wassenhove, V., and Ciuciu, P. (2018).
\newblock Self-similarity and multifractality in human brain activity: A
  wavelet-based analysis of scale-free brain dynamics.
\newblock {\em Journal of Neuroscience Methods}, 309:175 -- 187.

\bibitem[\protect\astroncite{Scherg and Von~Cramon}{1985}]{scherg-etal:85}
Scherg, M. and Von~Cramon, D. (1985).
\newblock Two bilateral sources of the late {AEP} as identified by a
  spatio-temporal dipole model.
\newblock {\em Electroencephalogr. Clin. Neurophysiol.}, 62(1):32--44.

\bibitem[\protect\astroncite{Stephen et~al.}{2013}]{stephen2013using}
Stephen, J.~M., Coffman, B.~A., Jung, R.~E., Bustillo, J.~R., Aine, C., and
  Calhoun, V.~D. (2013).
\newblock Using joint {ICA} to link function and structure using {MEG} and
  {DTI} in schizophrenia.
\newblock {\em Neuroimage}, 83:418--430.

\bibitem[\protect\astroncite{Subasi and Gursoy}{2010}]{subasi2010eeg}
Subasi, A. and Gursoy, M.~I. (2010).
\newblock {EEG} signal classification using {PCA}, {ICA}, {LDA} and support
  vector machines.
\newblock {\em Expert systems with applications}, 37(12):8659--8666.

\bibitem[\protect\astroncite{Tang et~al.}{2005}]{tang2005validation}
Tang, A.~C., Sutherland, M.~T., and McKinney, C.~J. (2005).
\newblock {Validation of SOBI components from high-density EEG}.
\newblock {\em NeuroImage}, 25(2):539--553.

\bibitem[\protect\astroncite{Urig{\"u}en and
  Garcia-Zapirain}{2015}]{uriguen2015eeg}
Urig{\"u}en, J.~A. and Garcia-Zapirain, B. (2015).
\newblock {EEG} artifact removal—state-of-the-art and guidelines.
\newblock {\em Journal of neural engineering}, 12(3):031001.

\bibitem[\protect\astroncite{Varoquaux et~al.}{2010}]{VAROQUAUX2010288}
Varoquaux, G., Sadaghiani, S., Pinel, P., Kleinschmidt, A., Poline, J., and
  Thirion, B. (2010).
\newblock A group model for stable multi-subject {ICA} on {fMRI} datasets.
\newblock {\em NeuroImage}, 51(1):288 -- 299.

\bibitem[\protect\astroncite{Vig{\'a}rio et~al.}{2000}]{vigario2000independent}
Vig{\'a}rio, R., Sarela, J., Jousmiki, V., Hamalainen, M., and Oja, E. (2000).
\newblock Independent component approach to the analysis of {EEG} and {MEG}
  recordings.
\newblock {\em IEEE transactions on biomedical engineering}, 47(5):589--593.

\bibitem[\protect\astroncite{Ziehe et~al.}{2004}]{ziehe2004fast}
Ziehe, A., Laskov, P., Nolte, G., and M\"{u}ller, K.-R. (2004).
\newblock A fast algorithm for joint diagonalization with non-orthogonal
  transformations and its application to blind source separation.
\newblock {\em Journal of Machine Learning Research}, 5(Jul):777--800.

\end{thebibliography}

\end{document}